\documentclass[8pt,prd,superscriptaddress,nofootinbib,notitlepage]{revtex4-1}

\usepackage{slashed}
\usepackage{caption}
\usepackage{subfigure}
\usepackage{amsmath,amssymb,graphicx,xcolor,mathtools}
\usepackage[colorlinks=true,linktocpage=green,linkcolor=blue,citecolor=blue]{hyperref}
\numberwithin{equation}{section}
\allowdisplaybreaks
\usepackage{hyperref}
\hypersetup{
	colorlinks=true,
	citecolor = blue,
	linkcolor= blue,
	filecolor= blue,      
	urlcolor= blue,
	pdftitle={HDL}
}
\usepackage{float}
\usepackage{graphics}
\usepackage{mathrsfs}
\usepackage{simpler-wick}
\usepackage{dsfont}
\usepackage{mathtools}

\DeclareMathOperator{\Tr}{Tr}

\def\be {\begin{equation}}
	\def\ee {\end{equation}}
\def\bea {\begin{eqnarray}}
	\def\eea {\end{eqnarray}}
\def\bc {\begin{center}}
	\def\ec {\end{center}}
\def\nn {\nonumber}

\DeclareMathAlphabet{\mathpzc}{OT1}{pzc}{m}{it}

\def\sumintof{\sum\!\!\!\!\!\!\!\!\!\int\limits_{\{P\}}}

\def\sumintoff{\sum\!\!\!\!\!\!\!\!\!\int\limits_{\{p_{0}\}}}

\def\sumintob{\sum\!\!\!\!\!\!\!\!\!\int\limits_{P}}
\def\sumintof{\sum\!\!\!\!\!\!\!\!\!\int\limits_{\{P\}}}
\def\sumintofk{\sum\!\!\!\!\!\!\!\!\!\int\limits_{\{K\}}}

\def\nn{\nonumber\\}

\begin{document}
	
	\title{One-loop HDL thermodynamics of a strongly magnetized isospin asymmetric cold quark matter }

	\author{Salman Ahamad Khan}
	\email{salmankhan.dx786@gmail.com}
	\affiliation{Department of Physics, Integral University, Lucknow - 226026, India
	}

	\author{Sarthak Satapathy}
	\email{sarthaks680@gmail.com}
	\affiliation{ Buxi Jagabandhu Bidyadhar Autonomous College,
Lewis Road, BJB Nagar, Bhubaneswar, Odisha 751014
	}
	\affiliation{School of Physical Sciences, National Institute of Science Education and Research,
An OCC of Homi Bhabha National Institute, Jatni-752050, India}
	
	\author{Sumit}
	\email{sumit@ph.iitr.ac.in}
\affiliation{School of Physics, Beijing Institute of Technology, Beijing 102488, China}

\begin{abstract}
We have computed the longitudinal pressure and magnetization of strongly magnetized cold QCD matter in the presence of both quark and isospin chemical potentials using the hard-dense-loop perturbation theory (HDLpt). For that purpose, we have first obtained the resummed quark and gluon propagators in the presence of a strong magnetic field and isospin density. We have found that pressure gets monotonically enhanced with both chemical potentials. Magnetization is found to be positive, which indicates the paramagnetic nature of the cold quark matter. We also discuss the resulting pressure anisotropy, where the transverse pressure is suppressed relative to the longitudinal pressure in the strong-field regime.
\end{abstract}

\maketitle

\section{Introduction}
\label{sec:intro}
Hadronic matter becomes deconfined in the plasma of quarks and gluons at extreme temperatures and/or densities. The equation of state (EoS) of this strongly interacting QCD matter is of great phenomenological importance across many areas of particle and astrophysics. The motivation to study the EoS of QCD matter also stems from the experimental programs at the Relativistic Heavy Ion Collider (RHIC) at Brookhaven and the Large Hadron Collider (LHC) at CERN, where the hot, dense phase of QCD is produced. The EoS is also used to study the various aspects of the QCD phase diagram and hydrodynamics of strongly interacting matter. The thermodynamic functions of hot QCD matter have been computed using the first-principle nonperturbative lattice approach by various groups~\cite{Bazavov:2017dus, Bazavov:2017dsy}. Since the lattice approach is not applicable at high densities due to the sign problem, efforts have been made to include the low-density region using various approximations~\cite{Borsanyi:2012cr, Guenther:2017hnx, Bazavov:2017dus, DElia:2016jqh}. Analytically, hard thermal loop perturbation theory (HTLpt) has been rigorously employed to obtain the thermodynamics of strongly interacting matter at high temperature and moderate density~\cite{Haque:2012my, Andersen:2010ct, Andersen:2010wu,  Haque:2014rua}.  In addition, HTL resummation has also been used to investigate medium-modified quark self-energies and dispersion relations beyond leading order~\cite{Sumit:2022bor}.  
\par
Recently, the existence of very high-density quark matter has been reported in the core of neutron stars~\cite{Annala:2019puf}, which motivates the precise knowledge of the EoS of strongly interacting matter at high densities and zero temperature. Such a matter is also known as cold quark matter. The first attempt to compute the thermodynamics of cold quark matter was made long ago~\cite{Freedman:1977gz, Freedman:1976ub, Baluni:1977ms, Toimela:1984xy}. Later, various efforts have been made to improve the earlier results using the perturbative QCD approach~\cite{Fraga:2004gz, Kurkela:2009gj, Blaizot:2000fc, Fraga:2013qra, Kurkela:2014vha, Annala:2017llu, Gorda:2018gpy, Gorda:2021kme}. Additionally, a very strong magnetic field is generated in non-central heavy-ion collisions, which motivates including it in the picture. A background magnetic field is also present during the cosmological phase transitions in the early universe and in the core of the neutron stars~\cite{Adhikari:2024bfa}. Hence, the EoS has been computed at finite magnetic field using first principle lattice approach at zero~\cite{Bonati:2013vba, Levkova:2013qda, Bali:2014kia} as well as at finite baryon density~\cite{Astrakhantsev:2024mat}. Similarly, the speed of sound in magnetized, dense nuclear matter has been explored within effective QCD models such as the Walecka model \cite{Mondal:2023baz} and the Nambu-Jona-Lasinio model \cite{Mondal:2024eaw}. The equation of state has also been computed in the ambiance of a magnetic field using the HTL perturbation theory~\cite{Rath:2017fdv, Bandyopadhyay:2017cle, Karmakar:2019tdp, Fraga:2023cef, Fraga:2023lzn}. It is imperative to study the thermal properties of cold and dense matter in the presence of very strong magnetic fields for a better understanding of the mass-radius relation of the compact stars since EoS enters as an input parameter~\cite{Kaspi:2017fwg}. In compact stars, the strength of the magnetic fields can be of the order of $10^{15}$ Gauss at the surface~\cite{Schaffner:Compact} and possibly much higher in the core (up to $10^{20}$ Gauss)~\cite{Ferrer:2010wz}. Experimental programs {\em like} NICER~\cite{Riley:2019yda, Miller:2019cac}, PSR J0740+6620~\cite{Riley:2021pdl, Miller:2021qha}, and the LIGO-VIRGO detection of gravitational waves from binary neutron star mergers~\cite{LIGOScientific:2018cki, LIGOScientific:2018hze}, give realistic estimates of the equation of state that describes quark stars and neutron stars~\cite{MUSES:2023hyz}. Recently, Podo and Santoni studied the properties of relativistic fermionic systems at finite density in magnetic fields using the path integral formulation~\cite{Podo:2023ute}. Given the importance of the background magnetic field on the EoS, we have recently computed the thermodynamic functions of strongly magnetized cold quark matter using the HDLpt and found that a strong magnetic field has a profound impact on the EoS. The pressure becomes anisotropic due to the Landau quantization of quark dynamics; in particular, dimensional reduction from 3 + 1 to 1 + 1 dimensions occurs, leading to anisotropic pressure. In all the above-mentioned studies, only the baryon chemical potential was considered. There are many systems in nature, such as the early universe, compact stars, and heavy-ion collisions, where the isospin chemical potential is nonzero.
{The effects of the isospin chemical potential on the properties of hot and dense quark matter have been studied extensively in absence~\cite{Son:PRL86'2001,Son:PAN64'2001, Toublan:PLB212'2003, Kogut:PRD70'2004,toublan:PLB'605'2005, Andersen:PRD75;2007, Cea:PRD;852012, Fraga:PRD79'2009, Andersen:JPG37'2009,kamikado:PLB718'2013, Ueda:PRD88'2013, Stiele:PLB729'2014,
Kanazawa:JHEP10'2014, Andersen:PRD93'2016} as well as in the presence of the magnetic field~\cite{Endrodi:PRD90'2014,prabal, yang}. Most of these studies have been performed either using first-principles lattice QCD or in the framework of various effective models.} One-loop thermodynamics of the cold quark matter has been computed using the  HDLpt by Andersen et al. \cite{Andersen:2002jz} to study the mass-radius relations of the dense quark stars. Recently, Fujimoto {\it et al.}  explored the effects of the nonzero quark mass on the thermodynamics of dense quark matter~\cite{Fujimoto:2020tjc}. Recently, we have also studied the  thermodynamics of strongly magnetized  cold quark matter using the HDL perturbation theory~\cite{Satapathy:2025jjx}.
  
\par
{{However, the simultaneous presence of a strong magnetic field, finite quark chemical potential, and finite isospin chemical potential at zero temperature has not been addressed within an HDL-resummed framework. The present work fills this gap by deriving the flavor-dependent quark and gluon contributions to the one-loop free energy of strongly magnetized cold quark matter with $\mu_{u,d} = \mu_q \pm \mu_I/2.$ It allows us to quantify the combined effects of Landau quantization, density-induced HDL screening, and isospin imbalance on the pressure, magnetization, and pressure anisotropy of cold quark matter.}} In the current study, we have computed the one-loop pressure of strongly magnetized cold quark matter using HDLpt in the presence of both quark and isospin chemical potentials. HDLpt is a resummation technique at high densities, similar to HTLpt, that accounts for screening and density-dependent damping effects~\cite{Manuel:1995td}.  

\par
The paper is organized as follows. In Sec. \ref{SE-Q}, we revisit the quark and gluon self-energy tensors in the presence of a strong magnetic field and isospin chemical potentials. In Sec. \ref{tp}, we have computed the quark and gluon contribution to the free energy. The results for pressure and magnetization are presented in Sec. \ref{Res}. We conclude in Sec. \ref{conc}. 
%%%%%%%%%%%%%%%%%%%%%%%%%%%%%%%%%%%%%%%%%
%%%%%%%%%%%%%%%%%%%%%%%%%%%%%%%%%%%%%%%%%
\section{Quark and gluon self-energy tensors in the presence of a magnetic field and finite isospin density}
\label{SE-Q}
In this section, we will derive the expression of the quark and gluon self-energies in the presence of a strong magnetic field at finite quark chemical potential $(\mu_q)$ and isospin chemical potential $(\mu_I)$ at vanishing temperature $(T)$. The strength of the magnetic field is determined by the number of Landau levels available to the system, and in our case, we consider the lowest Landau level (LLL), which corresponds to the strong magnetic field regime. 
%%%%%%%%%%%%%%%%%%%%%%%%%%%%%%%%%%%%%%%%%%%%%%%%%
\subsection{ Quark self-energy in the presence of a strong magnetic field and isospin density}
The quark self-energy at high chemical potential can be studied using the HDLpt, which is the high-density counterpart of the hard thermal loop perturbation theory (HTLpt). At high $T$ and strong magnetic field, the hierarchy of scales followed for the perturbation theory is $gT < T < \sqrt{q_f B}$, where $g$ is the coupling, $T$ is the temperature, $q_f$ is the charge of the fermion, and $B$ is the magnetic field. However, at finite $\mu$ and strong magnetic field, the hierarchy of scales assumes a slightly different structure. To deduce this hierarchy of scales, one has to look at the Landau quantization of energy in magnetic field, which is given by $E_l = \sqrt{p_3^2 + m^2 + 2l|q_f B|}$, where the Landau level $l \in [0, \infty)$, $p_3$ is the spatial momentum along the $z$ direction and $m$ is the mass of the fermion. The Fermi-Dirac distribution function $n_F(p_3, l, B, T, \mu)$ in the $T/\mu\to 0$ limit, 
\bea
 \displaystyle{\lim_{\frac{T}{\mu} \to 0}} n_F(p_3, l, B, T, \mu) =  \displaystyle{\lim_{\frac{T}{\mu} \to 0}} \frac{1}{  e^{(E_l-\mu)/T} + 1 } = \Theta(\mu - E_l), 
\eea
i.e.,  it assumes the profile of a Heaviside step function. 
{The LLL approximation is valid when the higher Landau levels are energetically suppressed. For a given flavor $f$, this requires
\begin{equation}
	l_{\rm max}^{(f)} = \left\lfloor \frac{\mu_f^2-m_f^2}{2|q_fB|} \right\rfloor \simeq 0, 
\end{equation}
or, in the massless limit,
\begin{equation}
	\mu_f^2 < 2|q_fB|.
\end{equation}
The HDL hierarchy additionally requires the external momentum to be soft compared with the hard scale set by the quark chemical potential, namely $g\mu_f\ll \mu_f$. Therefore, the regime of applicability of the present calculation is
\begin{equation}
	g\mu_f \ll \mu_f \lesssim \sqrt{2|q_fB|}.
\end{equation}
} To proceed with our calculation, we will first derive the general structure of the quark self-energy $\Sigma(p_0, p_3)$ in a strong magnetic field \cite{Karmakar:2019tdp}, at finite $T$ and $\mu$, and then take the $T/\mu \to 0$ limit of the obtained expressions. The general structure of $\Sigma(p_0, p_3)$ in terms of form factors $a, b, c, d$ is given by 
\bea
\Sigma(p_0, p_3) = a(p_0, p_3) \slashed{u} +  b(p_0, p_3) \slashed{n} + c(p_0, p_3) \gamma_5\slashed{u} + d(p_0, p_3)\gamma_5\slashed{n}, 
\label{seq-1} 
\eea  
\begin{figure}[h]
	\centering
	\includegraphics[scale = 0.6]{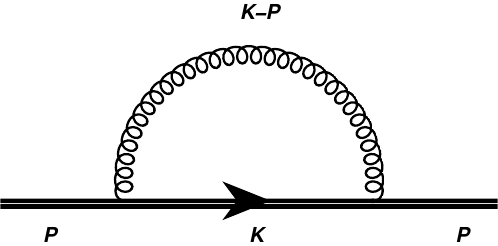}
	\caption{Feynman diagram for quark self-energy in the presence of a strong magnetic field (one loop).}
	\label{fig-FSE}
\end{figure}
%%%%%%%%%%%%%%%%%%%%%%%%%%%%%%%%%%%%%%%%%%%%%%%%%%%%%%%%%%%%%%%%%%%%%%%%%%%%%%%%%%
where $\slashed{f} = \gamma^\mu f_\mu$ for any 4-vector $f^\mu$, $u^\mu = (1,0,0,0)$ is the velocity of the heat bath, $n^\mu = (0,0,0,1)$ is the direction of the magnetic field, which is in the $z$-direction. The form factors are calculated as 
\bea
&& a = \frac{1}{4}\Tr[\Sigma \slashed{u}],~~b = \frac{1}{4}\Tr[\Sigma \slashed{n}],~~ c = \frac{1}{4}\Tr[\gamma_5\Sigma \slashed{u}],~~ d = \frac{1}{4}\Tr[\gamma_5\Sigma \slashed{n}] .
\eea  
%%%%%%%%%%%%%%%%%%%%%%%%%%%%%%%%%%%%%%%%%%%%%%%%%%%%%%%%%%%%%%%%
In Fig.~\ref{fig-FSE}, we show the one-loop quark self-energy Feynman diagram, where the doubled solid lines indicate a modification of the fermion propagator in a magnetic field. In the presence of a magnetic field, due to Landau quantization of the energy, the translational symmetry of the system is broken. This fact is exhibited in the coordinate space propagator of the fermion, where a multiplicative factor in the form of the Schwinger phase  \cite{Miransky:2015ava, Ghosh:2024hbf, Hattori:2023egw} is responsible for breaking the translational symmetry. It restricts a straightforward Fourier transform of the coordinate-space propagator to its momentum-space version. To treat this issue, such that we can work with the translationally invariant momentum parts of the propagator, we start by writing the quark self-energy in coordinate space $\Sigma(x,y)$ as 
%%%%%%%%%%%%%%%%%%%%%%%%%%%%%%%%%%%%%%%%%%%%%%%%%
\bea
\Sigma(x,y) = -ig^2 \gamma^\mu S(x,y) \gamma_\mu \Delta(x - y),
\eea 
where $g$ is the coupling, $S(x,y)$ is the coordinate space quark propagator with a broken translational symmetry, and $\Delta(x-y)$ is the scalar part of the gluon propagator. The expression of $S(x,y)$ with a Schwinger phase factor $\exp(i\Phi(x,y))$ is given by,
\bea
S(x,y) = e^{i\Phi(x,y)} S(x-y), 
\eea  
where $$\Phi(x,y) = q_f \int_y^x d\xi^\mu \bigg[ A_\mu + \frac{1}{2}F_{\mu\nu}(\xi - y)^\nu \bigg], $$ and $S(x-y)$ is the translationally invariant part of the fermion propagator. Physically measurable quantities should be independent of $\exp(i\Phi(x,y))$. Therefore, one would be able to gauge away this factor in the loop calculations. One thing to note is that the integrand in the Schwinger phase, i.e.,  $ A_\mu + \frac{1}{2}F_{\mu\nu}(\xi - y)^\nu \equiv f_\mu(\xi;y)$ is curl-free, which implies the path independence of the integration. If we consider a straight-line path parameterized as 
%%%%%%%%%%%%%%%%%%%%%%%%%%%%%%%%%%%%%%%%%%%%%%%%%
\bea
\xi^\mu = y^\mu + \sigma(x - y)^\mu ,~~~\sigma \in [0,1],
\eea 
and using the antisymmetric property of $F^{\mu\nu}$ we get 
\bea
\Phi(x,y) = q_f \int_0^1 d\sigma A_\mu (x - y)^\mu .  
\eea 
If the magnetic field is in the $z$ direction and we consider the symmetric gauge, i.e., $A_\mu(x) = \frac{B}{2}(0,-y,x,0)$, then a gauge transformation can be performed as
\bea
A_\mu(\xi) \to A_\mu'(\xi) = A_\mu + \frac{\partial }{\partial\xi^\mu} \Lambda(\xi), 
\eea  
where 
\bea
\Lambda(\xi) = \frac{B}{2} (y'\xi_1 - x'\xi_2).
\eea 
Following the above
\bea
\frac{\partial}{\partial \xi^\mu}\Lambda(\xi) = \frac{B}{2}(0, y', -x', 0), 
\eea 
due to which $\Phi(x,y)$ vanishes and the broken translational symmetry is recovered. Now the fermion propagator can be written as a Fourier transform as, 
\bea
S_F(x-y) = \int \frac{d^4k}{(2\pi)^4} e^{-ik\cdot(x-y)}S(k),
\eea 
where $S(k)$ is the momentum space propagator given by 
\bea
\hspace{-0.75cm}iS(k) = \int_0^\infty ds \exp\bigg[ is(k_\parallel^2 - m_f^2 - \frac{k_\perp^2}{q_fBs} \tan(q_fBs))  \bigg] \bigg[(\slashed{k}_\parallel + m)(1 + \gamma^1\gamma^2 \tan(q_fBs)) - \slashed{k}_\perp (1 + \tan^2(q_f Bs)) \bigg].
\eea
The fermion propagator above can be expressed as a sum over the Landau levels as 
\bea
iS(k) =  ie^{-\frac{k_\perp^2}{q_fB}}  \sum_{l = 0}^\infty  \frac{ (-1)^l D_l(q_fB, k) }{k_\parallel^2 - m_f^2 + i\epsilon}
\eea 
where $l \in [0, \infty)$. Here $D_l(q_fB, k)$ is given by 
\bea
D_l(q_fB, k) = (\slashed{k}_\parallel + m)\bigg[ (1 - i\gamma^1 \gamma^2) L_l\bigg(\frac{2k_\perp^2}{q_fB}\bigg) - (1 + i\gamma^1 \gamma^2 ) L_{l-1}\bigg(\frac{2k_\perp^2}{q_fB}\bigg)  \bigg] - 4 \slashed{k}_\perp L_{l-1}^1\bigg(\frac{2k_\perp^2}{q_fB}\bigg),
\eea 
where $L_l^\alpha(x)$ is the generalized Laguerre polynomial given by 
\bea
(1 - z)^{-(1+ \alpha)} \exp\bigg(\frac{xz}{z - 1}\bigg) = \sum_{l = 0}^\infty  L_l^\alpha(x) z^l. 
\eea 
In a strong magnetic field, all the fermions are confined to $l = 0$. In LLL, the Laguerre polynomials take the following values 
\bea
L_{-1}^1\bigg( \frac{2k_\perp^2}{q_fB} \bigg) = L_{-1}^0\bigg( \frac{2k_\perp^2}{q_fB} \bigg) = 0 \text{~and~} L_0\bigg( \frac{2k_\perp^2}{q_fB} \bigg) = 1.
\eea 
Therefore, the fermion propagator at $l = 0$ is given by
\bea
iS_F(k) = i e^{-\frac{k_\perp^2}{q_fB}}\frac{\slashed{k}_\parallel + m}{k_\parallel^2 - m^2} (1 - i\gamma_1\gamma_2).
\eea 
%%%%%%%%%%%%%%%%%%%%%%%%%%%%%%%%%%%%%%%%%%%%%%%%%%%%%%%%%%%%%%%%%%%%%%%%%%%%%%%  
The quark self-energy can now be written as
\bea
\Sigma(P) = -i g^2 C_F\int\frac{d^4k}{(2\pi)^4} \gamma_\mu S_F(K) \gamma^\mu \Delta(K-P),
\label{seq-3}
\eea 
where $C_F = (N_c^2 - 1)/(2N_c)$, $N_c$ is the number of colors, and $\Delta(K-P)$ is the  gluon propagator, which is not affected by the magnetic field 
\bea
\hspace{-1cm}\Delta(K-P) = \frac{1}{ (k_0 - p_0)^2 - (k - p)^2  } = \frac{1}{ (K-P)_\parallel^2 - (k-p)_\perp^2  }.
\label{seq-4}
\eea  
In Eq. (\ref{seq-4}), the four-momentum has been written in terms of the longitudinal and transverse parts as $K^\mu = K^\mu_\parallel + K^\mu_\perp$, where $K^\mu_\parallel = (k_0, 0, 0, k_3)$ and $K^\mu_\perp = (0, k_1, k_2, 0)$. With the help of the quark and gluon propagators, the quark self-energy in Eq. (\ref{seq-3}) becomes
\bea
\Sigma(P) 
%&=& -ig^2C_F \sum_f\int \frac{d^4K}{(2\pi)^4} e^{-k_\perp^2 / q_fB} \gamma_\mu \frac{\big( \slashed{K}_\parallel  - m_f \big)}{K_\parallel^2 - m_f^2} (\mathds{1} - i\gamma_1\gamma_2) \gamma^\mu \Delta(K-P)  \nn 
%%%%%%%%%%%%%%%%%%%%%%%%%%%%%%%%%%%%%%%%%%%%%%%%%%%%%%%%%%%%%%%%%%%%%%%%%%%%%%%%%%%%%%%%%%%
&=& -ig^2C_F \sum_f\int \frac{d^4K}{(2\pi)^4} e^{-k_\perp^2 / q_fB} \gamma_\mu \slashed{K}_\parallel  (\mathds{1} - i\gamma_1\gamma_2) \widetilde{\Delta}_\parallel(K) \Delta(K-P). 
\label{seq-6}
\eea 
We have neglected the current quark masses in comparison to $T, \mu$ and $\sqrt{q_fB}$. The subscript $``f"$ stands for the quark flavor and $\widetilde{\Delta}_\parallel(K) = (1)/(k_0^2 - k_3^2)$. In the evaluation of Eq. (\ref{seq-6}), the integration over four momentum can be written as a sum over the Matsubara frequencies and integration over the transverse and longitudinal components of the momentum as 
\bea
\int\frac{d^4K}{(2\pi)^4} \to iT \sum_{\{k_0\}}\int\frac{d^3k}{(2\pi)^3} \to iT \sum_{\{k_0\}}\int\frac{dk_3}{2\pi}\int\frac{d^2k_\perp}{(2\pi)^2}.
\eea 
Using Eq. (\ref{seq-6}) the form factors are given by 
\bea
a(p_0, p_3) &=& \frac{1}{4}\Tr[\Sigma \slashed{u}]  =  -\frac{g^2 C_F}{2}\sumintofk e^{-k_\perp^2/q_fB}\Tr\big[ \big(\mathds{1} + i\gamma_1\gamma_2\big) \slashed{K}_\parallel \slashed{u} \big]  \widetilde{\Delta}_\parallel(K)\Delta(K-P) \nn 
&=&  -2g^2C_F \sumintofk e^{-k_\perp^2/q_fB} k_0 \widetilde{\Delta}_\parallel(K)\Delta(K-P),
\label{seq-7a}
\eea 
%%%%%%%%%%%%%%%%%%%%%%%%%%%%%%%%%%%%%%%%%%%%%%%%%%%%%%%%%%%%%%%%%%%%
\bea
b(p_0, p_3) &=& \frac{1}{4}\Tr[\Sigma \slashed{n}] = \frac{g^2 C_F}{2}\sumintofk e^{-k_\perp^2/q_fB}\Tr\big[ \big(\mathds{1} + i\gamma_1\gamma_2\big) \slashed{K}_\parallel \slashed{n} \big]  \widetilde{\Delta}_\parallel(K)\Delta(K-P) \nn 
&=&  2g^2C_F \sumintofk e^{-k_\perp^2/q_fB} k_3 \widetilde{\Delta}_\parallel(K)\Delta(K-P),
\label{seq-7b}
\eea 
%%%%%%%%%%%%%%%%%%%%%%%%%%%%%%%%%%%%%%%%%%%%%%%%%%%%%%%%%%%%%%%%%%%%
\bea
c(p_0, p_3) &=& \frac{1}{4}\Tr[\gamma_5\Sigma \slashed{u}] =  -\frac{g^2 C_F}{2}\sumintofk e^{-k_\perp^2/q_fB}\Tr\big[ \gamma_5 \big(\mathds{1} + i\gamma_1\gamma_2\big) \slashed{K}_\parallel \slashed{u} \big]  \widetilde{\Delta}_\parallel(K)\Delta(K-P) \nn 
&=&  -2g^2C_F \sumintofk e^{-k_\perp^2/q_fB} k_3 \widetilde{\Delta}_\parallel(K)\Delta(K-P),
\label{seq-7c}
\eea
%%%%%%%%%%%%%%%%%%%%%%%%%%%%%%%%%%%%%%%%%%%%%%%%%%%%%%%%%%%%%%%%%%%%
\bea
d(p_0, p_3) &=& \frac{1}{4}\Tr[\gamma_5\Sigma \slashed{n}] =  \frac{g^2 C_F}{2}\sumintofk e^{-k_\perp^2/q_fB}\Tr\big[ \gamma_5 \big(\mathds{1} + i\gamma_1\gamma_2\big) \slashed{K}_\parallel \slashed{n} \big]  \widetilde{\Delta}_\parallel(K)\Delta(K-P) \nn 
&=&  2g^2C_F \sumintofk e^{-k_\perp^2/q_fB} k_0 \widetilde{\Delta}_\parallel(K)\Delta(K-P). 
\label{seq-7d}
\eea 
From Eqs. (\ref{seq-7a})-(\ref{seq-7d}), we notice that $a = -d$ and $b = -c$. To compute these form factors, we use the HTL approximation. To obtain the high-density limit of these form factors, we take the $T \to 0$ limit of $a(p_0, p_3)$ and $b(p_0, p_3)$. 
%In a strong magnetic field, the transverse component of the momentum is considered to be weaker in magnitude than the longitudinal component. 
We also use the approximation $|K-P|_\perp < |K - P|_\parallel$ to simplify our calculations. This would allow a geometric series expansion. Finally, we can write $\Delta(K-P)$ as  
\bea
\Delta(K-P) &=& \frac{1}{ \big(K -P\big)_\parallel^2 - \big(k - p\big)_\perp^2  }  
= \frac{1}{\big( K - P \big)_\parallel^2}\bigg[ 1 - \frac{(k - p)_\perp^2}{ (K - P)_\parallel^2 }   \bigg]^{-1}  \nn 
&\approx& \Delta_\parallel(K - P)  +   (k - p)_\perp^2 \Delta_\parallel^2(K-P)  . 
\label{seq-7e}
\eea 
After employing the above approximations, $a(p_0, p_3)$ can be written as
\bea
a(p_0, p_3) &=& \frac{1}{4}\Tr[\Sigma \slashed{u}] 
= -2g^2 C_F \int\frac{d^3k}{(2\pi)^3} e^{-k_\perp^2/ q_fB}\big[ \mathcal{N}_1 + (k-p)_\perp^2 \mathcal{N}_2 \big] , 
\label{seq-8}
\eea  
where $\mathcal{N}_1$ and $\mathcal{N}_2$ are given by
\bea
 \mathcal{N}_1 = T\sum\frac{k_0}{(K_\parallel^2)(K-P)_\parallel^2},~~~ 
\mathcal{N}_2 = T\sum\frac{k_0}{K_\parallel^2(K-P)_\parallel^4} = -\frac{1}{2k_3}\frac{\partial \mathcal{N}_1}{\partial p_3}.
\eea
On performing the transverse momentum integration in Eq. (\ref{seq-8}), $a(p_0, p_3)$ comes out to be
\bea
a = -g^2 C_F \frac{|q_fB|}{4\pi^2} \int_{-\infty}^{+\infty} dk_3 \big[\mathcal{N}_1 + \big(p_\perp^2 + q_fB\big)\mathcal{N}_2  \big] .
\label{a}
\eea  
On performing the Matsubara frequency sums \cite{Laine:2016hma,Haque:2024gva}, $\mathcal{N}_1$ and $\mathcal{N}_2$ are given by
\bea
\mathcal{N}_1 = -\frac{1}{4k_3}\bigg[\frac{n_B(k_3) + n_F(k_3 - \mu)}{p_0 + p_3} + \frac{n_B(k_3) + n_F(k_3 + \mu)}{p_0 - p_3} \bigg], ~~~
 \mathcal{N}_2 = -\frac{1}{2k_3}\frac{\partial \mathcal{N}_1}{\partial p_3} . 
\label{N1N2}
\eea 
In the dense limit, i.e., $T \to 0$, the Bose-Einstein distribution function $n_B(k_3)$ does not contribute. For the 2-flavor system considered in this work, we consider $u$ and $d$ quarks with $\mu_u$ and $\mu_d$ as their respective chemical potentials. The isospin chemical potential $\mu_I$ and  the quark chemical potential $\mu_q$ are related to the chemical potentials of $``u"$ and $``d"$ quarks through the relation 
\bea
\mu_I = \mu_u - \mu_d  \quad \quad  \mu_q = \frac{\mu_u + \mu_d}{2}.
 \eea
 These, in turn, help us obtain the relations for $\mu_u$ and $\mu_d$ respectively as 
\bea
\mu_{u} = \mu_q + \frac{\mu_I}{2} \, \quad \quad \quad \mu_{d} = \mu_q - \frac{\mu_I}{2},
\label{ud}
\eea  
which are then substituted into the calculations of the form factors and, subsequently, into those of the thermodynamic quantities. From Eq. (\ref{ud}), we find $\mu_u$ and $\mu_d$ are different. Therefore, the form factors need to be calculated separately for the $u$ and $d$ quarks. {As we consider the particle contribution only, the antiparticle distribution function $n_F(k_3 + \mu)$ in the $T/\mu \to 0$ limit does not contribute to the form factor calculations. }Integrating $\mathcal{N}_1$ we get
\bea
\displaystyle{\lim_{T/\mu_u \to 0} }  \int_{-\infty}^{+\infty} dk_3 \mathcal{N}_1 &=& -\displaystyle{\lim_{T/\mu_u \to 0}} \bigg( \frac{e^{\gamma_E}\Lambda^2}{4\pi} \bigg)^\epsilon \int_{-\infty}^{+\infty}\frac{dk_3}{4k_3} k_3^{-2\epsilon} \frac{ n_F(k_3 - \mu_q - \frac{\mu_I}{2}) }{p_0 + p_3}  \nn 
&=& - \bigg( \frac{e^{\gamma_E}\Lambda^2}{4\pi} \bigg)^\epsilon \int_{-\infty}^{+\infty}\frac{dk_3}{4k_3} k_3^{-2\epsilon} \frac{ \Theta(  \mu_q + \frac{\mu_I}{2} - |k_3|) }{p_0 + p_3}, 
\label{N_1}
\eea     
where $2\epsilon = 1- d$, $d$ refers to the number of spatial dimensions and $\big(\frac{e^{\gamma_E} \Lambda^2}{4\pi}\big)^\epsilon$ to the regularization factor which regulates the UV divergences. Moreover, $\Lambda$ is an arbitrary momentum scale that is, in general, a function of $T$ and $\mu$. Carrying out the integration, we get
\bea
\displaystyle{\lim_{T/\mu_u \to 0}}\int_{-\infty}^{+\infty}dk_3\mathcal{N}_1 = \frac{(p_0 - p_3)}{8P_\parallel^2}\bigg[ \frac{1}{\epsilon} + \bigg\{  \log\bigg(\frac{e^{\gamma_E } \Lambda^2}{\pi (\mu_I + 2\mu_q)^2}\bigg)\bigg\}\bigg\{1 +  \frac{  \epsilon   }{2}\log\bigg(\frac{e^{\gamma_E } \Lambda^2}{\pi (\mu_I + 2\mu_q)^2}\bigg)\bigg\}  \bigg] + \mathcal{O}(\epsilon^2).
\label{N1u}
\eea 
Using Eq. (\ref{N1u}) we can also calculate the integration involving $\mathcal{N}_2$ as 
\bea
\displaystyle{\lim_{T/\mu_u \to 0}} \int_{-\infty}^{+\infty}dk_3 |q_fB| \mathcal{N}_2 = \frac{|q_fB|(p_0 - p_3)^2}{4P_\parallel^4(\mu_I + 2\mu_q)}\bigg[ 1 + \epsilon\bigg\{-2 + \log\bigg(  \frac{e^{\gamma_E} \Lambda^2}{\pi(\mu_I + 2\mu_q)^2}  \bigg)\bigg\}  \bigg]. 
\label{N2u}
\eea 
Substituting Eqs. (\ref{N1u}) and (\ref{N2u}) into the expression of $a(p_0, p_3)$, we get the desired expression for the $u$ quark, which we call $a_u(p_0, p_3)$. Neglecting terms upto $\mathcal{O}(\epsilon)$ we get 
\bea
\hspace{-0.55cm}a_u(p_0, p_3) = -g^2C_F\frac{|q_fB|}{4\pi^2}\bigg[  \frac{1}{8(p_0 + p_3)\epsilon}  +  \frac{(p_0 - p_3)}{8P_\parallel^2}\bigg\{  \log\bigg(  \frac{e^{\gamma_E  }\Lambda^2}{\pi (\mu_I + 2\mu_q)^2} \bigg)    + \frac{2|q_fB| }{P_\parallel^2(\mu_I + 2\mu_q)}(p_0 - p_3)     \bigg\} \bigg] . 
\label{au}
\eea 
For the $d$ quark, the corresponding $a(p_0, p_3)$ term will be called $a_d(p_0, p_3)$, which will depend on whether $\mu_q - \frac{\mu_I}{2}$ is greater than or less than 0, which in turn decides the nature of the distribution function. For $\mu_q - \frac{\mu_I}{2} > 0$, the $n_F(k_3 - \mu)$ term in the expression of $\mathcal{N}_1$ in Eq. (\ref{N1N2}) contributes, whereas for $\mu_q - \frac{\mu_I}{2} < 0$, the $n_F(k_3 + \mu)$ term contributes. Following the computations similar to those for the $u$ quark, we get 
\bea
\hspace{-0.55cm}a_{d,\pm}(p_0, p_3) = -g^2C_F\frac{|q_fB|}{4\pi^2}\bigg[  \mp\frac{1}{8(p_0 \pm p_3)\epsilon} + \frac{(p_0 \mp p_3)}{8P_\parallel^2}\bigg\{  \log\bigg(   \frac{e^{\gamma_E}\Lambda^2}{\pi (2\mu_q - \mu_I)^2}\bigg)  \mp \frac{2|q_fB|}{P_\parallel^2(\mu_I - 2\mu_q)}(p_0 \mp p_3)      \bigg\} \bigg] ,
\label{ad}
\eea  
where the subscript $\pm$ in $a_{d,\pm}$ denotes whether $\mu_q > \mu_I/2$ or $<\mu_I/2$. The calculation of $b(p_0, p_3)$ for $u$ and $d$ quarks is obtained using the expression
\bea
b(p_0, p_3) = g^2C_F\frac{|q_fB|}{4\pi^2}\int_{-\infty}^{+\infty} dk_3~k_3\big[\mathcal{N}_3 + |q_fB|\mathcal{N}_4\big] ,
\label{b}
\eea 
which has been obtained after performing the Matsubara frequency sums. Here $\mathcal{N}_3$ and $\mathcal{N}_4$ are given by 
\bea
\mathcal{N}_3 = \frac{1}{4k_3^2}\bigg[ \frac{n_B(k_3) + n_F(k_3-\mu)}{p_0 + p_3} - \bigg\{\frac{n_B(k_3) + n_F(k_3 +\mu)}{p_0 - p_3}\bigg\}  \bigg],~~~\mathcal{N}_4 = -\frac{1}{2k_3}\frac{\partial  \mathcal{N}_3}{\partial p_3}.
\label{N3N4}
\eea 
Here we observe that both particle and antiparticle distribution functions are present in the expressions of $\mathcal{N}_3$ and $\mathcal{N}_4$. Following a similar calculational procedure as done for $a(p_0, p_3)$, we will do the same for $b(p_0, p_3)$. For the $u$ quark, the integration of $\mathcal{N}_3$ is given by 
\bea
\displaystyle{\lim_{T/\mu \to 0}} \int_{-\infty}^{+\infty}dk_3~ k_3 \mathcal{N}_3 =  -\frac{(p_0 - p_3)}{8P_\parallel^2}\bigg[\frac{1}{\epsilon} + \bigg\{\log\bigg(  \frac{e^{\gamma_E}  \Lambda^2 }{\pi(\mu_I + 2\mu_q)^2} \bigg)\bigg\} \bigg\{1 + \frac{\epsilon}{2} \log\bigg(  \frac{e^{\gamma_E}  \Lambda^2 }{\pi(\mu_I + 2\mu_q)^2} \bigg)   \bigg\}\bigg] + \mathcal{O}(\epsilon^2).
\label{N3u}
\eea 
Using Eq. (\ref{N3u}) we can also calculate the integration involving $\mathcal{N}_4$ as 
\bea
\displaystyle{\lim_{T/\mu \to 0}} \int_{-\infty}^{+\infty}dk_3|q_fB| k_3 \mathcal{N}_4 =  -\frac{|q_fB|(p_0 - p_3)^2}{4P_\parallel^4(\mu_I + 2\mu_q)}\bigg[1   +  \epsilon\bigg\{   -2 +  \log\bigg(   \frac{e^{\gamma_E} \Lambda^2}{ \pi (\mu_I + 2\mu_q)^2 } \bigg) \bigg\} \bigg] +\mathcal{O}(\epsilon^2).
\label{N4u}
\eea  
Substituting Eqs. (\ref{N3u}) and (\ref{N4u}) into the expression of $b(p_0, p_3)$ and neglecting the $\mathcal{O}(\epsilon)$ terms, we get the structure function for the $u$ quark which reads as follows
\bea
\hspace{-0.55cm}b_u(p_0, p_3) = -g^2C_F\frac{|q_fB|}{4\pi^2}\bigg[  \frac{1}{8(p_0 + p_3)\epsilon}  + \frac{(p_0 - p_3)}{8P_\parallel^2}\bigg\{   \log\bigg(   \frac{e^{\gamma_E}\Lambda^2}{\pi(\mu_I + 2\mu_q)^2} \bigg) + \frac{2|q_fB|}{P_\parallel^2(\mu_I + 2\mu_q)}(p_0 - p_3) \bigg\} \bigg] . 
\label{bu}
\eea  
However, for the $d$ quark, we again have to take into account whether $\mu_q$ is greater or less than $\mu_I/2$, with this information entering the distribution functions. The final expression of $b_{d,\pm}(p_0, p_3)$ is given by 
\bea
b_{d,\pm}(p_0,p_3) = -g^2C_F\frac{|q_fB|}{4\pi^2}\bigg[  \pm \frac{1}{8(p_0 + p_3)\epsilon} \pm \frac{(p_0 \mp p_3)}{8P_\parallel^2}\bigg\{  \log\bigg(   \frac{e^{\gamma_E}\Lambda^2}{\pi (2\mu_q -\mu_I)^2} \bigg)  \mp \frac{2|q_fB| }{P_\parallel^2(\mu_I - 2\mu_q)}  (p_0 \mp p_3)\bigg\}    \bigg].
\label{bd} 
\eea 
Using the expressions of $a_u, a_{d,\pm}, b_u$ and $b_{d,\pm}$, we can now calculate the free energy of quarks at finite isospin density in a strong magnetic field. 
%%%%%%%%%%%%%%%%%%%%%%%%%%%%%%%%%%%%%%%%%%%%%%%%%%%%%%%%%%%%%%%%%%%%%%%%%%%%%%%%%%%%%%%%%%%%%%%%%%%%%%%%%%%%%%%%%%%%%%%%%%%%%%%%%%%%%%%%%%%%%%%%%%%%%%
\subsection{Gluon self-energy in the presence of a strong magnetic field and isospin density}
\label{GSE}
A medium when subjected to a magnetic field suffers a breaking of the Lorentz and rotational symmetry. Keeping this in mind, the general structure of the gauge boson self-energy  can be written as~\cite{Karmakar:2018aig}
\begin{eqnarray}
\Pi^{\mu\nu}(P)=b(P)B^{\mu\nu}(P)+c(P)R^{\mu\nu}(P)+d(P)M^{\mu\nu}(P)
+a(P)N^{\mu\nu}(P),
\label{self_decomposition}
\end{eqnarray}
where the projection tensors in Eq. \eqref{self_decomposition} are constructed as
\bea
B^{\mu\nu}(P) = \frac{{\bar{u}}^\mu{\bar{u}}^\nu}{{\bar{u}}^2}, \quad \quad
R^{\mu\nu}(P) = g_{\perp}^{\mu\nu}-\frac{P_{\perp}^{\mu}P_{\perp}^{\nu}}
{P_{\perp}^2}, \quad \quad 
M^{\mu\nu}(P) = \frac{{\bar{n}}^\mu{\bar{n}}^\nu}{{\bar{n}}^2}, \quad \quad 
N^{\mu\nu}(P) = \frac{{\bar{u}}^\mu{\bar{n}}^\nu+{\bar{u}}^\nu{\bar{n}}^\mu}
{\sqrt{{\bar{u}}^2}\sqrt{{\bar{n}}^2}}.
\eea
Here $u^\mu=(1,0,0,0)$ is the four velocity of the heat bath and 
$n_\mu=(0,0,0,1)$ is the unit vector of magnetic field along the $z$ direction. Further we can construct ${\bar{u}}^\mu$ and ${\bar{n}}^\mu$  in terms of $u^\mu, g^{\mu\nu}$ and $P^\mu$ as  
\begin{eqnarray}
\bar{u}^\mu = \left(g^{\mu\nu}-\frac{P^\mu P^\nu}{P^2}\right)u_\nu,~~~
\bar{n}^\mu = \left(\tilde{g}^{\mu\nu}-\frac{\tilde{P}^\mu\tilde{P}^\nu}
{\tilde{P}^2}\right)n_\nu,
\end{eqnarray}
where ${\tilde{g}}^{\mu\nu}=g^{\mu\nu}-u^\mu u^\nu$ and 
$\tilde{P}^\mu=P^\mu-(P\cdot u)~u^\mu$. The form factors defined in  \eqref{self_decomposition} can be evaluated via contraction of the rank 2 tensors with $\Pi_{\mu\nu}$ as  
\bea
\hspace{-.5cm} b(P) = B^{\mu\nu}(P)~\Pi_{\mu\nu}(P), \quad
%\label{structure_b} 
c(P) = R^{\mu\nu}(P)~\Pi_{\mu\nu}(P),  \quad
%\label{structure_c}\\
d(P) = M^{\mu\nu}(P)~\Pi_{\mu\nu}(P), \quad 
%\label{structure_d}\\
a(P) = \frac{1}{2}N^{\mu\nu}(P)~\Pi_{\mu\nu}(P)
%\label{structure_a}.
\eea
Clearly, form factors in general depend on temperature and magnetic field. The gluon self-energy is a sum of the gluon loops  $\Pi^{\mu\nu}_g$ and the fermion loop $\Pi^{\mu\nu}_s$ 
\bea
\Pi^{\mu\nu} = \Pi^{\mu\nu}_g + \Pi^{\mu\nu}_s  ,
\eea 
%%%%%%%%%%%%%%%%%%%%%%%%%%%%%%%%%%%%%%%%%%%%%%%%%%%%%%%%%%%%%%%%%%%%%%%%%%%%%%%%%%
\begin{figure}[h]
	\centering
	\includegraphics[scale = 0.6]{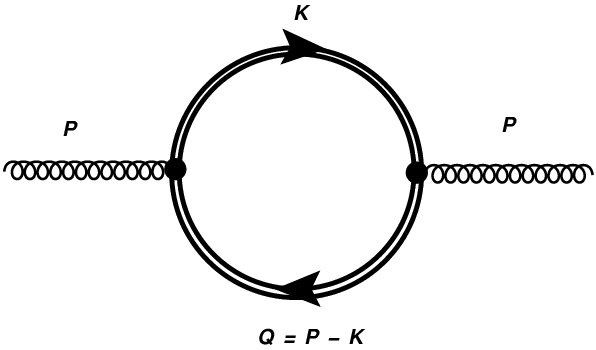}
	\caption{Quark loop contribution to the gluon self-energy in the strong magnetic field. Double straight lines show the modification in the quark propagator due to the magnetic field}.
	\label{fig-GSE}
\end{figure}
%%%%%%%%%%%%%%%%%%%%%%%%%%%%%%%%%%%%%%%%%%%%%%%%%%%%%%%%%%%%%%%%%%%%%%%%%%%%%%%%%%
In the $T/\mu_{u,d} \to 0$ limit, $\Pi^{\mu\nu}_g$ has a vanishing contribution. We will therefore consider only the contribution of $\Pi^{\mu\nu}_s$. From Ref.~\cite{Karmakar:2018aig}, the final expressions of the form factors can be written as 
\bea
a(p_0, p_3) =  \sum_f e^{-p_\perp^2/q_fB} \frac{\sqrt{\overline{n}^2}}{\sqrt{\overline{u}^2}} \bigg(\frac{g^2 |q_fB|}{4\pi^2} \bigg)\frac{p_0 p_3}{p_0^2 - p_3^2} \quad \quad
b(p_0, p_3) = -\sum_f \frac{g^2 |q_fB|}{4\pi^2 \overline{u}^2} e^{-p_\perp^2/2q_fB} \frac{p_3^2}{p_0^2 - p_3^2} \nn
c(p_0, p_3) = 0 \quad \quad 
d(p_0, p_3)= - b(p_0, p_3),
\label{FG-4}
\eea
where $\overline{n}^2 = -p_\perp^2 /p^2$ and $\overline{u}^2 = -p^2/P^2$.
These quark and gluon form factors will be employed in the calculation of the quark and gluon contributions to the free energy, respectively, in the next section.

%%%%%%%%%%%%%%%%%%%%%%%%%%%%%%%%%%%%%%%%%%%%%%%%%%%%%%%%%%%%%%%%%%%%%%%%%%%
%%%%%%%%%%%%%%%%%%%%%%%%%%%%%%%%%%%%%%%%%%%%%%%%%%%%%%%%%%%%%%%%%%%%%%%%%%%
\section{Pressure of the strongly magnetized  isospin asymmetric cold quark matter}
\label{tp}
The total free energy of the dense and cold quark matter can be written as the sum of the quark free energy and the gluon free energy as 
\bea
F=F_f+F_g
\eea
where $F_f$ and $F_g$ are the quark and gluon contributions to the free energy, respectively. In the subsequent subsections, we will compute the quark and gluon contributions to the free energy, respectively.
\subsection{Quark contribution to the free energy}
\label{FE}
%In this section, we present the calculation of the quark-free energy in a strong magnetic field using the form factors calculated in Sec. \ref{SE-Q}. Previously, this calculation was performed in Ref. \cite{Karmakar:2019tdp} by substituting the form factors $a(p_0, p_3)$ and $b(p_0, p_3)$ at finite temperature and density. Here, we briefly recapitulate those calculations and, to calculate the free energy, we substitute the form factors at zero temperature and finite density, as derived in Eq. (\ref{ab_final}). The free energy of the $f$ flavor quark can be written as 
In this section, we will compute the quark contribution to free energy in a strong magnetic field background, exploiting the form factors calculated in Sec. \ref{SE-Q}. For the strong magnetic field case, the free energy has already been calculated in a high-temperature scenario using the HTL perturbation theory~\cite{Karmakar:2019tdp}. The quark part of the free energy is given by
%\bea
%F_{f} = -N_c\sumintof \frac{d^3p}{(2\pi)^3} \ln\big(\det \big[ S_{\text{eff}}^{-1}(p_0, p_3) \big] \big).
%\label{fe-1}
%\eea 
\bea
F_{f} = -N_c\sumintof \ln\big(\det \big[ S_{\text{eff}}^{-1}(p_0, p_3) \big] \big).
\label{fe-1}
\eea 
where subscript $f$ corresponds to the quark flavours and $S_\text{eff}$ is the resummed quark propagator. The sum integral can be written as 
\bea
\sumintof \equiv T \sum_{\{p_0\}} \int \frac{d^3p}{(2\pi)^3}.
\eea 

%%%%%%%%%%%%%%%%%%%%%%%%%%%%%%%%%%%%%%%%%%%%%%%%%%%%%%%%%%%%%%%%%%%%%%%%%%%
%%%%%%%%%%%%%%%%%%%%%%%%%%%%%%%%%%%%%%%%%%%%%%%%%%%%%%%%%%%%%%%%%%%%%%%%%%%
Before the computation of the resummed quark propagator, we will first focus on the  form of the same in coordinate space, which can be written as
\bea
S_\text{eff}(u,u') = i \langle u|[ \gamma^\mu \Pi_\mu + \Sigma  ]^{-1}  |u' \rangle,  
\label{seff-1}
\eea     
where $u = (t,x,y,z)$ is the space co-ordinate,  $\Pi^\mu $ is the canonical momentum defined as $\Pi^0 = i\partial_t + \mu$ and $\Pi^k = i\partial^k + gA^k$ and $\Sigma$ refers to the quark self-energy. We have chosen the Landau gauge for the gauge field, i.e., $A^\mu = (0, 0, Bx, 0)$. Now the modified self-energy reads as 
\bea
\widetilde{\Sigma} = -\gamma^2 \Sigma \gamma^2 = -\Sigma. 
\label{seff-11} 
\eea 
The matrix element in Eq. (\ref{seff-1}) can be rewritten as 
\bea
S_\text{eff}(u,u') 
&=& i \langle u| [ \slashed{\Pi} +  \widetilde{\Sigma}  ]  [ \slashed{\Pi}^2 -  \Sigma^2 ]^{-1}  |u' \rangle  . 
\eea   
The Fourier transform of $S_\text{eff}(u,u')$ along the $t$ and $z$, directions can be written as 
\bea
S_\text{eff}(p_0, p_3; u_\perp, u_\perp') = \int dt~dz~e^{ip_0(t-t') - ip_3(z - z') }S_\text{eff}(u, u'),
\label{seff-2}
\eea
where 
\bea S_\text{eff}(p_0, p_3; u_\perp, u_\perp') = e^{i\Phi(u_\perp, u'_\perp)} S_\text{eff}(p_0, p_3; u_\perp - u_\perp'),~~~u_\perp = (x,y).
\label{seff-3}
\eea
We notice that the translational invariance is broken due to the Schwinger phase factor $e^{i\Phi(u_\perp, u'_\perp)}$, and $S_\text{eff}(p_0, p_3; u_\perp - u_\perp')$ is the translationally invariant part of the propagator. We can break  $\Pi^\mu$ into $\Pi^0$, $\Pi^\perp$ and $\Pi^3$, and in the lowest Landau level ($l = 0$), the eigenvalue of $\Pi^\perp = \sqrt{2l|q_fB|}|_{l = 0} = 0$. Thus, the propagator takes the form of  
\bea
S_\text{eff}(p_0, p_3; u_\perp, u_\perp') = \langle u_\perp | [ \gamma^0 p_0 - \gamma^3 p_3 + \widetilde{\Sigma}  ] [p_0^2 - p_z^2 - \Sigma^2 ]^{-1} |  u'_\perp \rangle.
\eea   
We can write the matrix element using the  spectral expansion of the unit operator as 
\bea
\int dp \langle u_\perp|p \rangle \langle r | u_\perp' \rangle = \int_{-\infty}^{+\infty} dp~ \psi_p(u_\perp) \psi^*_p(u_\perp') = \delta^2(u_\perp - u_\perp'), 
\label{spec-exp}
\eea 
where $\psi_p(u_\perp), \psi^*_p(u_\perp')$ refers to the normalized wave functions.
Thus $S_\text{eff}(p_0, p_3; u_\perp, u_\perp')$ can now be written as
\bea
S_\text{eff}(p_0, p_3; u_\perp, u_\perp')  &=&  \int_{-\infty}^{+\infty} dp~ \langle u_\perp | p \rangle   [ \gamma^0 p_0 - \gamma^3 p_3 + \widetilde{\Sigma}  ] [p_0^2 - p_3^2 - \Sigma^2 ]^{-1}  \langle p | u_\perp' \rangle  \nn 
%%%%%%%%%%%%%%%%%%%%%%%%%%%%%%%%%%%%%%%%%%%%%%%%%%%%%%%%%%%%%%%%%%%%%%%%%%%%%%%%%%%%%%%%%%%%%%%%%%%%%%%%
%%%%%%%%%%%%%%%%%%%%%%%%%%%%%%%%%%%%%%%%%%%%%%%%%%%%%%%%%%%%%%%%%%%%%%%%%%%%%%%%%%%%%%%%%%%%%%%%%%%%%%%%
&=&  \frac{q_fB}{2\pi}   \exp( i\Phi(u_\perp, u_\perp')  )  \exp\bigg( -q_fB\frac{(u_\perp - u_\perp')^2}{4}  \bigg)\bigg[ \frac{ (\gamma\cdot p)_\parallel + \widetilde{\Sigma} }{  (\gamma\cdot p)_\parallel^2 - \Sigma^2 }  \bigg ], 
\label{seff-4}  
\eea  
where 
\bea
\Phi(u_\perp, u'_\perp) = \frac{q_fB}{2}(x + x')(y - y').
\eea 
From Eq. (\ref{seff-3}), using the translationally invariant part of the propagator and doing a Fourier transform of Eq. (\ref{seff-4}) with respect to the transverse spatial coordinates,
\bea
S_\text{eff}(p_0, p_z) &=& \int d^2u_\perp~e^{-ip_\perp\cdot u_\perp}  S_\text{eff}(p_0, p_z; u_\perp ) \nn 
&=& \frac{q_fB}{2\pi}\int d^2u_\perp~\exp\bigg(-p_\perp\cdot u_\perp -q_fB\frac{u_\perp ^2}{4}  \bigg) \bigg[ \frac{ (\gamma\cdot p)_\parallel + \widetilde{\Sigma} }{  (\gamma\cdot p)_\parallel^2 - \Sigma^2 }  \bigg ] 
= 2e^{-\frac{p_\perp^2 }{ q_fB}}  \bigg[ \frac{ (\gamma\cdot p)_\parallel + \widetilde{\Sigma} }{  (\gamma\cdot p)_\parallel^2 - \Sigma^2 }  \bigg ]. 
\label{seff-5}
\eea   
%%%%%%%%%%%%%%%%%%%%%%%%%%%%%%%%%%%%%%%%%%%%%%%%%%%%%%%%%%%%%%%%%%%%%%%%%%%%%%%%%%%%%%%%%%%%%%%%%%%%%%%%
%%%%%%%%%%%%%%%%%%%%%%%%%%%%%%%%%%%%%%%%%%%%%%%%%%%%%%%%%%%%%%%%%%%%%%%%%%%%%%%%%%%%%%%%%%%%%%%%%%%%%%%%
Similarly, we can also write the inverse of the effective fermion propagator $S_\text{eff}^{-1}(p_0, p_3)$ in coordinate space in terms of matrix element (exploiting Eq. (\ref{seff-1}) and (\ref{seff-11})) as
\bea
S_\text{eff}^{-1}(u,u') = -i\langle u| \gamma^\mu \Pi_\mu + \Sigma   |u' \rangle.
\eea 
In momentum space, the effective propagator takes the form
\bea
S_\text{eff}^{-1}(p_0, p_3) = \gamma^0p_0 - \gamma^3p_3 + \Sigma(p_0,p_3) = (p_0 + a_f)\gamma^0 + (b_f - p_3)\gamma^3 + c\gamma_5\gamma^0 + d\gamma_5\gamma^3,   
\eea 
where the subscript $``f"$ in $a_f$ and $b_f$ denotes the flavour of the quark. 
%%%%%%%%%%%%%%%%%%%%%%%%%%%%%%%%%%%%%%%%%%%%%%%%%%%%%%%%%%%%%%%%%%%%%%%%%%%
%%%%%%%%%%%%%%%%%%%%%%%%%%%%%%%%%%%%%%%%%%%%%%%%%%%%%%%%%%%%%%%%%%%%%%%%%%%
%and $S_{\text{eff}}^{-1}(p_0, p_3)$ is the inverse of the effective quark propagator that reads as
%\bea
%S_{\text{eff}}^{-1}(p_0, p_3) &=& \slashed{P}_\parallel + \Sigma (p_0, p_3) = (p_0 + a)\slashed{u} + (b - p_3)\slashed{n} + c\gamma_5\slashed{u} + d\gamma_5\slashed{n} \nn
%&=& (p_0 + a)\gamma^0 + (b - p_3)\gamma^3 + c\gamma_5\gamma^0 + d\gamma_5\gamma^3.
%\label{fe-2}
%\eea 
%\bea
%S_{\text{eff}}^{-1}(p_0, p_3) &=& (p_0 + a)\gamma^0 + (b - p_3)\gamma^3 + c\gamma_5\gamma^0 + d\gamma_5\gamma^3.
%\label{fe-2}
%\eea 
%The determinant of the $S_{\text{eff}}^{-1}(p_0, p_3)$ is calculated as
%\bea
%\det \big[ S_{\text{eff}}^{-1}(p_0, p_3) \big] &=& ((b + c - p_3)^2 - (a + d + p_0)^2)((-b + c + p_3)^2 - (a - d + p_0)^2)  \nn 
%&=& P_\parallel^4\bigg[ 1 + \frac{4a^2 - 4b^2 + 4ap_0 + 4bp_3}{P_\parallel^2} \bigg].
%\label{fe-3}
%\eea 
The determinant of the $S_{\text{eff}}^{-1}(p_0, p_3)$ comes out to be
\bea
\det \big[ S_{\text{eff}}^{-1}(p_0, p_3) \big] &=& P_\parallel^4\bigg[ 1 + \frac{4a_f^2 - 4b_f^2 + 4a_fp_0 + 4b_fp_3}{P_\parallel^2} \bigg].
\label{fe-3}
\eea 
%Substituting the expression of the determinant in Eq. (\ref{fe-1}), we get 
With the help of the above  Eq.~\eqref{fe-3}, the free energy of quarks can be written as 
%\bea
%F_{f} &=& -2N_c\sumintof \frac{d^3p}{(2\pi)^3} \ln P_\parallel^2 - N_c\sumintof \frac{d^3p}{(2\pi)^3} \ln\bigg[ 1 + \frac{4a^2 - 4b^2 + 4ap_0 + 4bp_3}{P_\parallel^2}  \bigg] 
%\label{fe-4-0} \\
%&=& F_{0,f} + F'_{f} ,
%\label{fe-4}
%\eea 
\bea
F_{f} &=& -2N_c\sumintof \ln P_\parallel^2 - N_c\sumintof \ln\bigg[ 1 + \frac{4a_f^2 - 4b_f^2 + 4a_fp_0 + 4b_fp_3}{P_\parallel^2}  \bigg] = F_{0,f} + F^{\prime}_{f} ,
\label{fe-4}
\eea 
%where $F_{0,f}$ is the ideal free energy given by the first sum-integral in Eq. (\ref{fe-4-0}) and $F_{f}'$ is the one-loop correction to the free energy. An expansion of $F_{f}'$ up to $\mathcal{O}(g^4)$ terms reads
where $F_{0,f}$ and $F_{f}^{\prime}$ correspond to the ideal free energy and one-loop correction to the free energy, respectively.  An expansion of $F_{f}^{\prime}$ upto $\mathcal{O}(g^4)$ terms reads
\bea
F_{f}^{\prime} = -N_c\frac{|q_fB|}{(2\pi)^2}\sumintoff dp_3 \bigg[ \frac{4 (a_fp_0 + b_fp_3)}{P_\parallel^2} + \frac{4\big(a_f^2 P_\parallel^2 - b_f^2 P_\parallel^2 - 2a_f^2 p_0^2 - 2b_f^2 p_3^2 - 4a_fb_f p_0p_3\big)}{P_\parallel^4}  \bigg] + \mathcal{O}(g^6).
\label{fe-5}
\eea  
%where the above expansion is valid for $g^2\big(q_fB/\mu_f^2\big) < %1$, where $g << 1$ and $q_fB/\mu_f^2 \gtrsim 1$, a condition valid for strong magnetic field.
With the help of the form factors  $a_f(p_0, p_3)$ and $b_f(p_0, p_3)$ in $F_f'$ computed in the previous subsection, the free energy can be written as 
\bea
F_{f}^{\prime}(\mu, q_fB)  
%%%%%%%%%%%%%%%%%%%%%%%%%%%%%%%%%%%%%%%%%%%%%%%%%%%%%%%%%%%%%%%%%%%%%%%%%
&=& -N_c  \frac{|q_fB|}{(2\pi)^2}  \bigg[  -\frac{1}{8}\bigg(\frac{|q_fB| }{\mu_f}\bigg)^2 \bigg( g^2C_F\frac{|q_f B|}{4\pi^2} \bigg)^2\big\{ \mathcal{I}_{630} + \mathcal{I}_{603} - \mathcal{I}_{612} - \mathcal{I}_{621}  \big\} -\frac{1}{2} \bigg(g^2C_F\frac{|q_f B|}{4\pi^2} \bigg) \log\bigg(\frac{e^{\gamma_E}\mu_f^2}{\pi\Lambda^2}\bigg) \nn
%%%%%%%%%%%%%%%%%%%%%%%%%%%%%%%%%%%%%%%%%%%%%%%%%%%%%%%%%%%%%%%%%%%%%%%%%%%%%%%%%%%%%%%%%%%%%%%%%%
&& \times  \big\{  \mathcal{I}_{210} - \mathcal{I}_{201} \big\} -\frac{1}{8}  \bigg(g^2C_F\frac{|q_f B|}{4\pi^2} \bigg)^2 \bigg\{ \log\bigg(\frac{e^{\gamma_E}\mu_f^2}{\pi\Lambda_f^2}\bigg)  \bigg\}^2 \big\{ \mathcal{I}_{420} - 2\mathcal{I}_{411} + \mathcal{I}_{402} \big\}\bigg],
%%%%%%%%%%%%%%%%%%%%%%%%%%%%%%%%%%%%%%%%%%%%%%%%%%%%%%%%%%%%%%%%%%%%%%%%%
\label{I_abc}
\eea
where $\mathcal{I}_{\alpha\beta\gamma} \equiv \mathcal{I}_{\alpha\beta\gamma}(\mu_f)$ corresponds to the various dense sum integrals, which are computed by taking the $T \to 0$ limit of the sum integrals at finite $T$ and $\mu$. These dense sum integrals were first computed by Gorda et al. \cite{Gorda:2022yex}, where it was shown that the limits $T = 0$ and $T \to 0$ are not the same. In this work, we have derived a general structure of the sum integrals $\mathcal{I}_{\alpha\beta\gamma}$ (see the Appendix \ref{App-A}), which is given by
\bea  
\hspace{-0.75cm}\mathcal{I}_{\alpha\beta\omega}(\mu_f) &=&  \displaystyle{\lim_{T/\mu_f \to 0}}\sumintof  \frac{ p_0^{2\beta} p^{2\omega} }{ P^{2\alpha} }  \nn
%%%%%%%%%%%%%%%%%%%%%%%%%%%%%%%%%%%%%%%%%%%%%%%%%%%%%%%%%%%%%%%%%%%
&=& \bigg( \frac{e^{\gamma_E} \Lambda^2}{4\pi}  \bigg)^\epsilon \frac{i\mu_f}{2\pi}\frac{ \Gamma(\alpha - \omega - d/2) \Gamma(d/2 + \omega) (i\mu_f)^{d + 2\omega - 2\alpha + 2\beta} }{ (4\pi)^{d/2} \Gamma(\alpha) \Gamma(d/2) \big( 1 + d + 2\omega - 2\alpha + 2\beta  \big)  } \Big\{ \big( -1\big)^{d + 2\omega - 2\alpha + 2\beta} -1 \Big\}. 
\label{SI}
\eea 
In Eq. (\ref{SI}), $p_0 = ip_n = i(2n + 1)\pi T + \mu_f,$ $n\in Z$, is the discrete Matsubara frequency and $p$ is the  momentum. Following Eq. (\ref{SI}), the dense sum integrals have been calculated (see the Appendix \ref{App-A}). The ideal part of the free energy of the $f$ flavor quark is given by
\bea
F_{0,f}(\mu_f, q_fB) = -2N_c\sumintof\frac{d^3p}{(2\pi)^3}\ln P_{\parallel}^2  
=  -N_c\frac{|q_fB|}{4\pi^2}\mu_f^2 .
\label{F_Ideal}
\eea 
The quark-free energy carries a tree-level contribution \cite{Podo:2023ute} due to the presence of the background magnetic field, which does not carry any medium effects, i.e., $-B^2/2$. Adding this term, the expression of free energy in Eq. (\ref{fe-4}) can be rewritten as  
\bea
F_f(\mu_f, q_fB) = F_f^{\prime} + F_{0,f} - \frac{B^2}{2}.
\label{F_HDL}
\eea 
 With the help of the free energy, we can compute  longitudinal pressure $P_L$ and the transverse pressure $P_\perp$ as 
\bea
P_L = -F_f,~~P_\perp = - F_f - eB\cdot \mathcal{M},
\eea
where $\mathcal{M} = -\frac{\partial F_f}{\partial (eB)}$ is the magnetization per unit volume. 
%%%%%%%%%%%%%%%%%%%%%%%%%%%%%%%%%%%%%%%%%%%%%%%%%%%%%%%%%%%%%%%%%%%%%%%%%%%%%%%%%%%%%%%%%%%%%%%%%%
%%%%%%%%%%%%%%%%%%%%%%%%%%%%%%%%%%%%%%%%%%%%%%%%%%%%%%%%%%%%%%%%%%%%%%%%%%%%%%%%%%%%%%%%%%%%%%%%%%
%%%%%%%%%%%%%%%%%%%%%%%%%%%%%%%%%%%%%%%%%%%%%%%%%%%%%%%%%%%%%%%%%%%%%%%%%%%%%%%%%%%%%%%%%%%%%%%%%%
%%%%%%%%%%%%%%%%%%%%%%%%%%%%%%%%%%%%%%%%%%%%%%%%%%%%%%%%%%%%%%%%%%%%%%%%%%%%%%%%%%%%%%%%%%%%%%%%%%
\subsection{Gluon contribution to the free energy}
\label{FG}

%%%%%%%%%%%%%%%%%%%%%%%%%%%%%%%%%%%%%%%%%%%%%%%%%%%%%%%%%%%%%%%%%%%%%%%%%%%%%%%%%%

%%%%%%%%%%%%%%%%%%%%%%%%%%%%%%%%%%%%%%%%%%%%%%%%%%%%%%%%%%%%%%%%%%%%%%%%%%%%%%%%%%

%The free energy of gluons at finite temperature and strong magnetic field has been computed in Ref. \cite{Karmakar:2019tdp}, using the general expression of the gauge boson self-energy at finite temperature and magnetic field derived in Ref. \cite{Karmakar:2018aig}. 
 The gluon part of the free energy does not include the ideal part, since it vanishes in the $T \to 0$ limit. The medium-dependent part has been computed at finite temperature and density as~\cite{Karmakar:2019tdp} 
\bea
F_g' = -(N_c^2 - 1)\sumintob \bigg[ \frac{b + c + d}{2P^2} + \frac{b^2 + c^2 + d^2}{4P^4} \bigg].
\label{FG5}
\eea 
The form factors $a$, $b$, $c$, and $d$ have been computed in Eq.~(\ref{FG-4}). The first term in the above equation vanishes, while the second term gives
%\bea
%\frac{b + c + d}{P^2} &=& 0  
%\label{FG6-0}\\
%\frac{b^2 + c^2 + d^2}{4P^4} &=& \frac{1}{2}\sum_{f_1,f_2} \bigg(\frac{g^2 B}{2\pi}\bigg)^2 q_{f_1} q_{f_2} \sumintob  e^{-\frac{p_\perp^2}{2B}\big(\frac{1}{q_{f_1}} + \frac{1}{q_{f_2}}\big)} \frac{p_3^4}{p^4(p_0^2 - p_3^2)^2}. 
%\label{FG6} 
%\eea 
\bea
\frac{b + c + d}{P^2} &=& 0  \quad , \quad \quad 
\frac{b^2 + c^2 + d^2}{4P^4} = \frac{1}{2}\sum_{f_1,f_2} \bigg(\frac{g^2 B}{4\pi^{2}}\bigg)^2 q_{f_1} q_{f_2} \sumintob  e^{-\frac{p_\perp^2}{2B}\big(\frac{1}{q_{f_1}} + \frac{1}{q_{f_2}}\big)} \frac{p_3^4}{p^4(p_0^2 - p_3^2)^2}. 
\label{FG6} 
\eea 
The dense sum integral in Eq. (\ref{FG6}) can be computed as 
%\bea
%\displaystyle{\lim_{T \to 0}}\sumintob  \frac{p_3^4}{p^4(p_0^2 - p_3^2)^2} &=& \displaystyle{\lim_{T \to 0}}  -\bigg(\frac{e^{\gamma_E}\Lambda^2}{4\pi}\bigg)^\epsilon \int \frac{d^{3-2\epsilon}p}{(2\pi)^3} e^{-p_\perp^2/2B\big(\frac{1}{q_{f_1}} + \frac{1}{q_{f_2}}\big)} \frac{p_3}{2p^4}\bigg[ \beta \frac{\partial}{\partial \beta}  - 1\bigg] n_B(p_3)  \nn 
%&=& \frac{1}{(4\pi)^2}\bigg[ \frac{1}{2\epsilon} + \ln 4  + \gamma_E \bigg] + \text{higher orders in } \frac{1}{q_fB}, 
%\label{FG7}  
%\eea 
\bea
\displaystyle{\lim_{T \to 0}}\sumintob \,  e^{-\frac{p_\perp^2}{2B}\big(\frac{1}{q_{f_1}} + \frac{1}{q_{f_2}}\big)} \, \frac{p_3^4}{p^4(p_0^2 - p_3^2)^2} 
 &=& \displaystyle{\lim_{T \to 0}}  -\bigg(\frac{e^{\gamma_E}\Lambda^2}{4\pi}\bigg)^\epsilon \int \frac{d^{3-2\epsilon}p}{(2\pi)^3} e^{-p_\perp^2/2B\big(\frac{1}{q_{f_1}} + \frac{1}{q_{f_2}}\big)} \frac{p_3}{2p^4}\bigg[ \beta \frac{\partial}{\partial \beta}  - 1\bigg] n_B(p_3)  \nn 
&=& \frac{1}{(4\pi)^2}\bigg[ \frac{1}{2\epsilon} + \frac{\ln 4}{2}  + \gamma_E \bigg] + \text{higher orders in } \frac{1}{q_fB}, 
\label{FG7}  
\eea 
where $\beta = T^{-1}$. The gluon part of the free energy  $F_g^{\prime}$ is found to be
%\bea
%F_g^{\prime} &=& -\frac{N_c^2 - 1}{2}\sum_{f_1, f_2}\bigg(\frac{g^2 B}{2\pi}\bigg)^2 e^{-\frac{p_\perp^2}{2B}\big(\frac{1}{q_{f_1}} + \frac{1}{q_{f_2}}\big)}   \frac{1}{(4\pi)^2}\bigg[ \frac{1}{2\epsilon} + \ln 4  + \gamma_E \bigg] ,
%\label{FG8}
%\eea 
\bea
F_g^{\prime} &=& -\frac{(N_c^2 - 1)}{2}\sum_{f_1, f_2} q_{f_1} q_{f_2}\bigg(\frac{g^2 B}{4\pi^{2}}\bigg)^2 \frac{1}{(4\pi)^2}\bigg[ \frac{1}{2\epsilon} + \frac{\ln 4}{2}  + \gamma_E \bigg] ,
\label{FG8}
\eea 
which is divergent due to the presence of $\mathcal{O}(\epsilon^{-1})$ term.  We also notice that there is no medium dependence in $F_g^{\prime}$, which is the manifestation of the lowest Landau level dynamics in the strong magnetic field. {Concerning the divergent part of the free energy, we note that the divergence is a purely vacuum contribution. Therefore, this divergence can be taken care of with the help of a vacuum energy density renormalization counterterm \cite{Andersen:1999fw}. To implement this renormalization, we add a counterterm free energy $F_g^\text{ct}$ given by}
%\bea
%F_g^\text{ct} = \frac{N_c^2 - 1}{4\epsilon}\sum_{f_1, f_2}\bigg(\frac{g^2 B}{2\pi}\bigg)^2 e^{-\frac{p_\perp^2}{2B}\big(\frac{1}{q_{f_1}} + \frac{1}{q_{f_2}}\big)}   \frac{1}{(4\pi)^2},
%\label{FG9}  
%\eea 
\bea
F_g^\text{ct} = \frac{(N_c^2 - 1)}{4\epsilon}\sum_{f_1, f_2} q_{f_1} q_{f_2} \, \bigg(\frac{g^2 B}{4\pi^{2}}\bigg)^2 \frac{1}{(4\pi)^2},
\label{FG9}  
\eea 
to $F_g$, which renormalizes the free energy. Thus, the final expression of the gluon free energy $F_g$ is given by 
\bea
F_g = F_g^{\prime} + F_g^\text{ct}.
\label{FG10}
\eea 
%%%%%%%%%%%%%%%%%%%%%%%%%%%%%%%%%%%%%%%%%%%%%%%%%%%%%%%%%%%%%%%%%%%%%%%%%%%%%%%%%%%%%%%%%%%%%%%%%%

\section{Results and Discussions}

The expression of the one-loop running coupling constant in a strong magnetic field \cite{Ayala:2018wux} is given by 
\bea
\alpha_s(\Lambda^2, |q_fB|) = \frac{ \alpha_s(\Lambda^2)  }{1 + b_1\alpha_s(\Lambda^2)\ln\big( \frac{\Lambda^2}{\Lambda^2 + |q_fB|} \big)    },~~\text{where~~} \alpha_s(\Lambda^2) = \frac{1}{ b_1\ln\big( \Lambda^2 / \Lambda^2_{ \overline{\text{MS}} } \big)   },
\label{res-1}
\eea    
where $b_1 = (11N_c - 2N_f)/12\pi$, $\Lambda_{ \overline{ \text{MS} } } = 176$ MeV \cite{Karmakar:2020mnj}. {The renormalization scale $\Lambda$ for our work has been considered in the limit of low temperature, which is $\displaystyle{\lim_{T/\mu\to 0}}\Lambda(T,\mu) = \displaystyle{\lim_{T/\mu\to 0}}2\pi \sqrt{T^2 + \mu^2/\pi^2} = 2\mu$ }. Since we are interested only in the thermomagnetic corrections, we have dropped the vacuum contribution $-B^2/2$ in our results.
\label{Res}
\begin{figure}[H]
	\includegraphics[scale = 0.375]{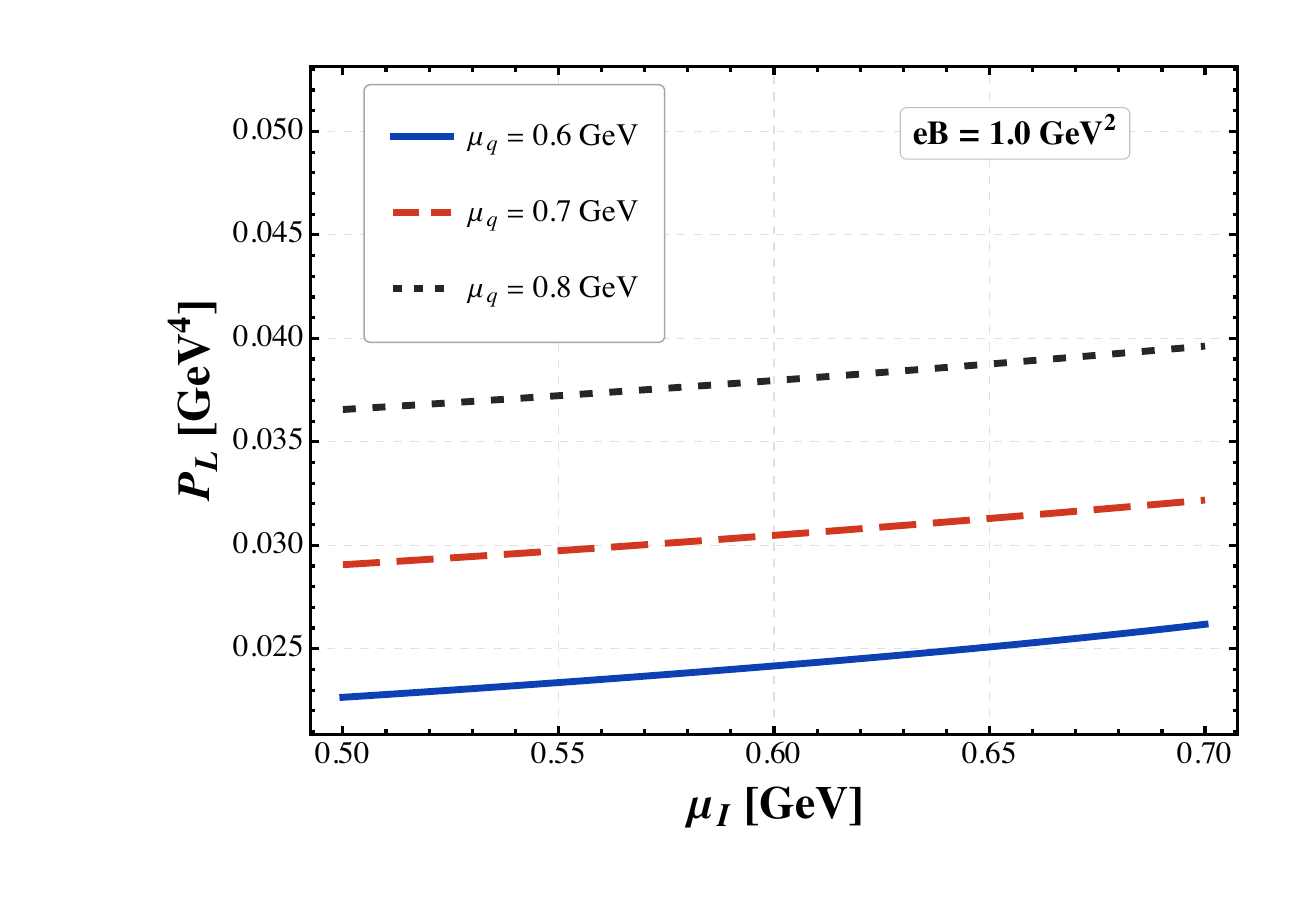}
	\includegraphics[scale = 0.375]{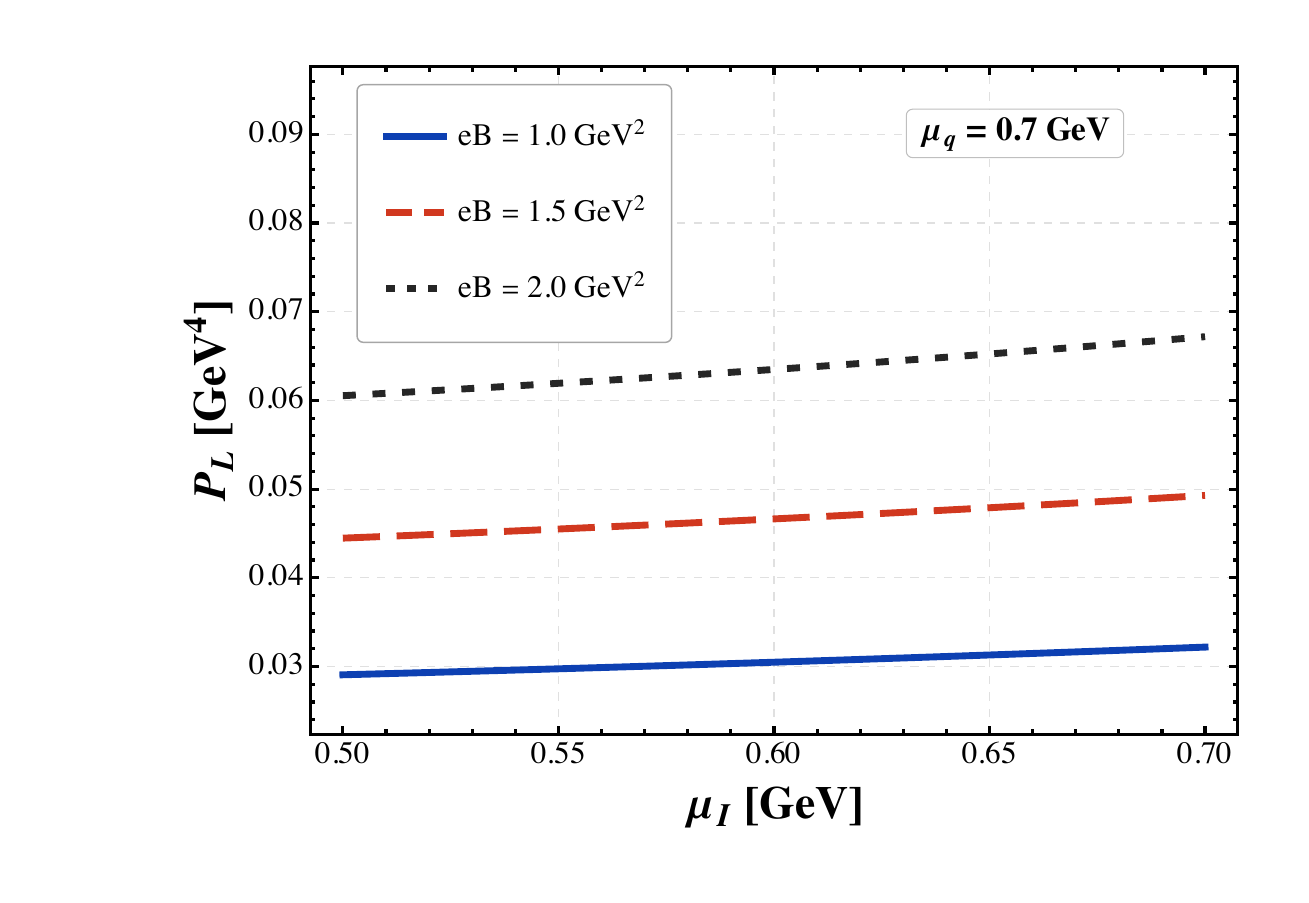} 
\caption{Variation of longitudinal pressure with isospin chemical potential at different values of the quark chemical potentials at $|eB| = 1.0$ GeV$^2$  (left panel). Variation of the longitudinal pressure with isospin chemical potential at different strengths of magnetic field at $\mu_q = 0.7$ GeV (right panel). The plots have been obtained for $\Lambda = 2\mu$.}
	\label{Pressure}
\end{figure}

\begin{figure}[H]
	\includegraphics[scale = 0.375]{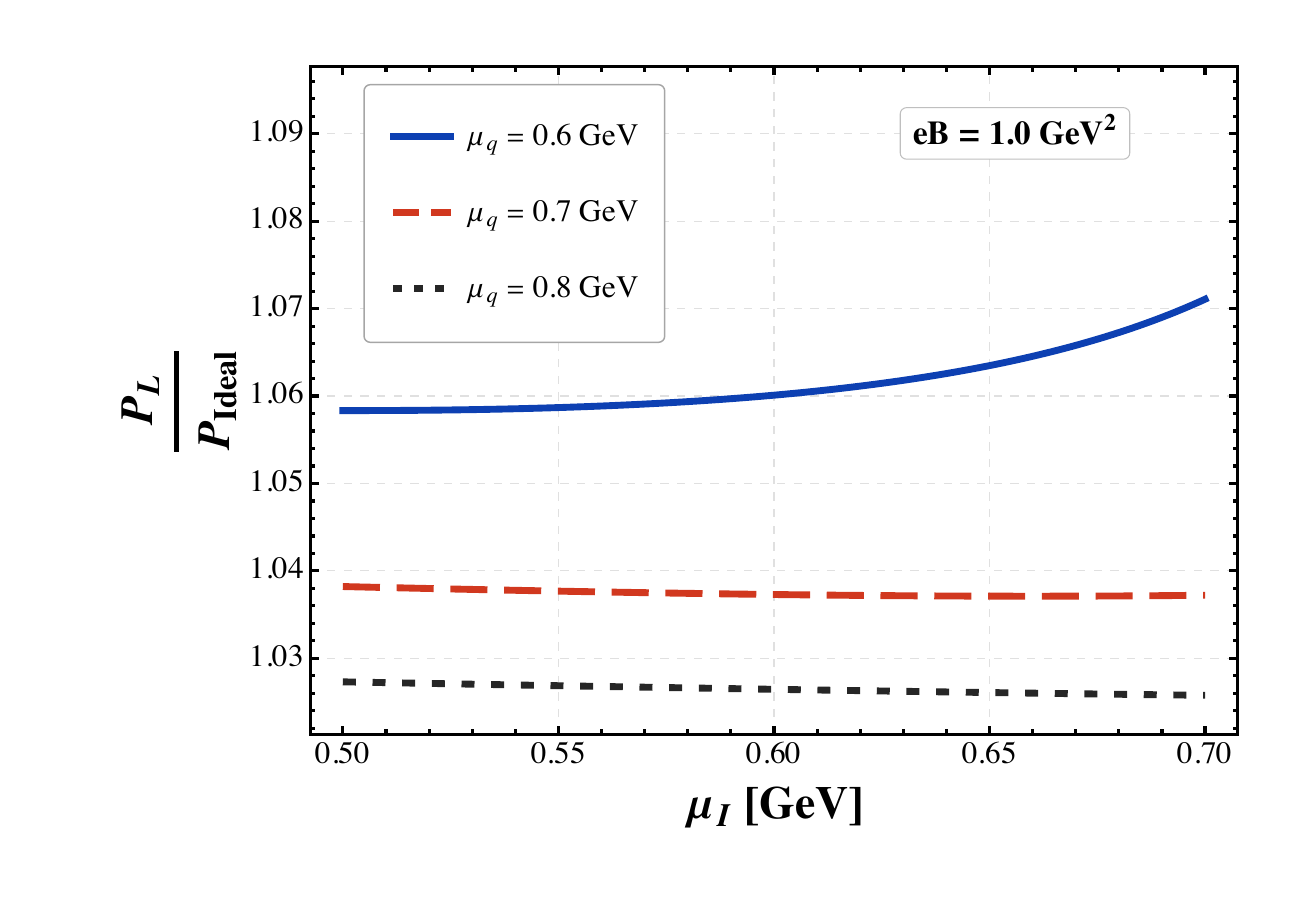}
	\includegraphics[scale = 0.375]{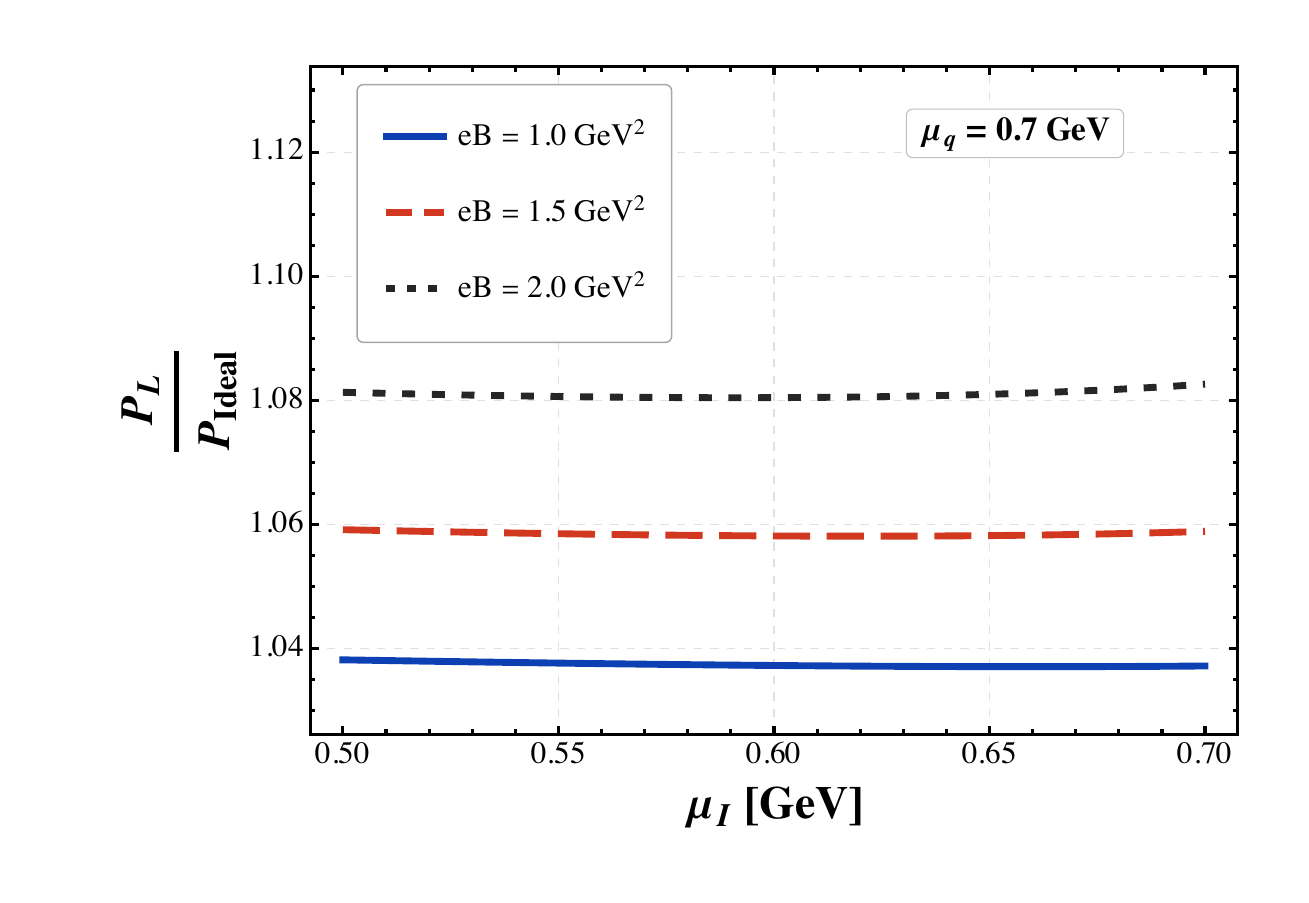}
	\caption{Variation of ratio of HDL pressure to the ideal pressure with isospin chemical potential at different values of the quark chemical potentials at $|eB| = 1.0$ GeV$^2$ (left panel). Variation of the ratio of HDL pressure to the ideal pressure with isospin chemical potential at different strengths of magnetic field at $\mu_q = 0.7$ GeV (right panel). The plots have been obtained for $\Lambda = 2\mu$.}
	\label{pressure_ratio}
\end{figure}
\begin{figure}[H]
	\includegraphics[scale = 0.375]{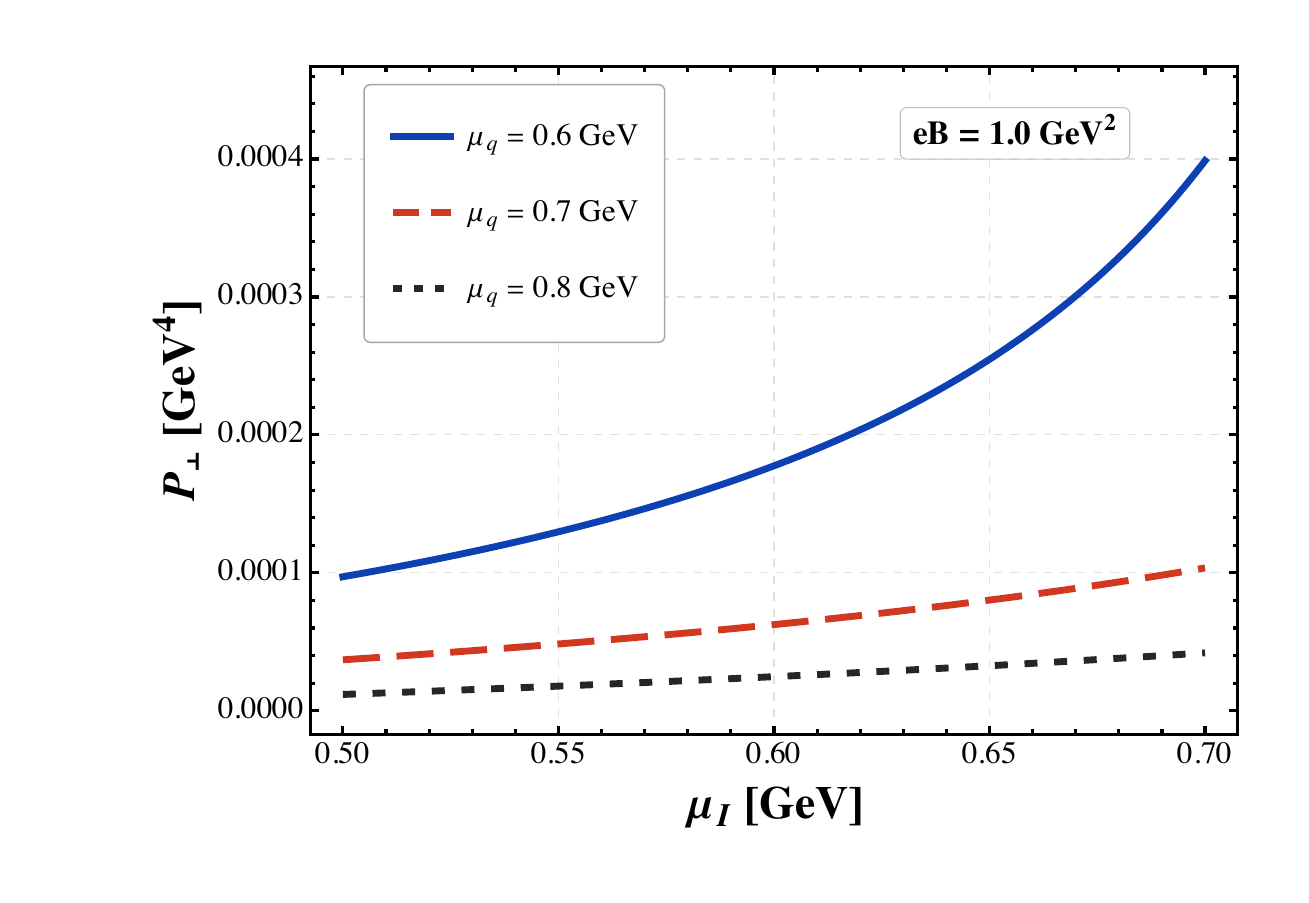}
	\includegraphics[scale = 0.375]{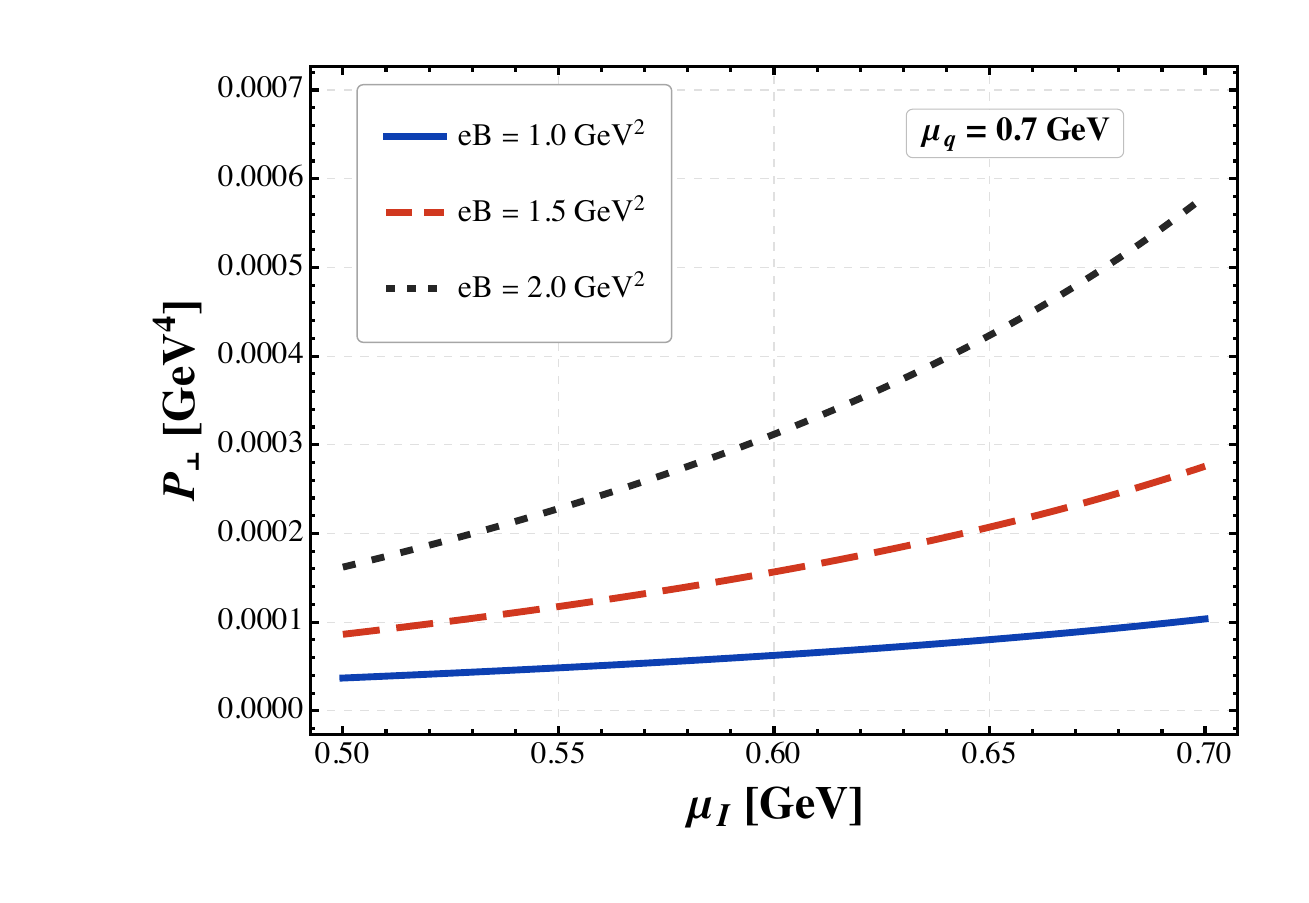} 
	%\caption{Variation of the transverse pressure with isospin chemical potential at different values of the quark chemical potentials at $|eB| = 1.0$ GeV$^2$ (left panel). Variation of the magnetization with isospin chemical potential at different strengths of magnetic field at $\mu_q = 0.7$ GeV (right panel). The plots have been obtained for $\Lambda = 2\mu$. }
    \caption{Variation of the transverse pressure with isospin chemical potential at different values of the quark chemical potentials at $|eB|=1.0~{\rm GeV}^2$ (left panel). Variation of the transverse pressure with isospin chemical potential at different strengths of magnetic field at $\mu_q=0.7~{\rm GeV}$ (right panel). The plots have been obtained for $\Lambda=2\mu$.}
	\label{P_T}
\end{figure}

 In the left panel of figure~\ref{Pressure}, we have displayed the longitudinal pressure of cold quark matter as a function of the quark and isospin chemical potentials, respectively. {We have considered $\mu_q$ and $\mu_I$ in the following range $$0.6~\text{GeV~}\le \mu_q\le 0.8~\text{GeV}~\text{and}~0.5~\text{GeV~}\le \mu_I\le 0.7~\text{GeV},$$ For these ranges of $\mu_q$ and $\mu_I$, $\mu_u$ and $\mu_d$ fall in the range of $$0.85~\text{GeV~}\le \mu_u\le 1.15~\text{GeV} \text{~and~} 0.25~\text{GeV~}\le \mu_d\le 0.55~\text{GeV},$$ which satisfy the LLL condition for the dense case and are subsequently used in the numerical calculations. The magnetic field range has been considered between 1 GeV$^2$ and 2 GeV$^2$.} From the plot, we notice that the pressure increases with both $\mu_q$ and $\mu_I$. Similarly, in the right panel of figure \ref{Pressure}, we plot pressure as a function of the isospin chemical potential and the magnetic field. We observe a similar increase in the magnetic field as well. 
{The monotonic increase of the pressure with $\mu_q$ is a direct consequence of the increase in Fermi momentum and the corresponding enhancement of the degeneracy pressure. The dependence on $\mu_I$ is flavor-asymmetric: a positive isospin chemical potential increases $\mu_u$ and decreases $\mu_d$. Since the $u$ quark carries a larger electric charge, its larger Landau degeneracy factor dominates in the parameter range considered. The increase in $eB$ originates from the magnetic field-dependent density of states in the LLL approximation. To quantify the effects of interactions on pressure, we also show the ratio of the pressure calculated using the HDLpt to the ideal pressure in Figure~\ref{pressure_ratio}. HDLpt uses resummed quark and gluon propagators that systematically incorporate interaction effects. The ratio is found to be greater than one for the whole range of values of $\mu_q$, $\mu_I$, and $eB$, which indicates an increase of pressure in the presence of interactions. The ratio approaches unity as both chemical potentials increase. This is a well-known behavior, as cold quark matter approaches an ideal system due to the weakening of interactions at asymptotic freedom in QCD. The positive magnetization indicates the paramagnetic nature of the medium, and the results, when compared with Ref. \cite{Ferraris:2025fva}, show a similar behavior. This behavior can be attributed to the fact that in the LLL-dominated regime, the spin-polarized quark states aligned with the magnetic field dominate over the diamagnetic orbital contribution, leading to positive magnetization. This behavior is consistent with the expected paramagnetic nature of deconfined quark matter in a strong magnetic background. The positive magnetization can also be attributed to the phenomenon of paramagnetic squeezing \cite{Bali:2013owa}, which occurs because of the high-pressure anisotropy in a strong magnetic field, i.e., $P_L >> P_\perp$, leading to the compression of the matter along the magnetic field.} {Concerning the transverse pressure, since the external magnetic field selects a preferred spatial direction, the pressure becomes anisotropic. The longitudinal pressure is obtained from the free energy as
\begin{equation}
	P_L=-F,
\end{equation}
whereas the transverse pressure is
\begin{equation}
	P_\perp=-F-eB\cdot\mathcal{M}=P_L-eB\cdot\mathcal{M}.
\end{equation}
The positive magnetization obtained in this work, therefore, implies
\begin{equation}
	P_\perp<P_L.
\end{equation}
It shows that the magnetic field suppresses the transverse pressure relative to the longitudinal one. In the LLL regime, the transverse motion of quarks is strongly restricted, and the longitudinal dynamics dominate the thermodynamics. The transverse pressure is fixed once $P_L$ and $\mathcal{M}$ are known, and hence the pressure anisotropy is directly controlled by the magnetization. From Figure \ref{P_T} we observe that $P_\perp$ has a suppressed magnitude as compared to $P_L$. The behavior of the plots is similar to what has been obtained in Ref. \cite{Ferraris:2025fva}. }

{{Note that, following Eq.~(\ref{ud}) and the hierarchy of energy scales discussed before, the chemical potentials $\mu_q$ and $\mu_I$ are chosen in such a way that the LLL condition is satisfied. We emphasize that the numerical results presented in this work are restricted to the normal quark-matter regime where both effective flavor chemical potentials remain positive, i.e., $\mu_u, \mu_d > 0$. Evidently, for the $d$ quark, the condition $\mu_q > \mu_I/2$ is a necessity. The regime $\mu_q < \mu_I/2$, in which the effective $ d$-quark chemical potential changes sign and pion-condensed phases may become relevant, lies outside the domain of validity of the present HDLpt framework. A proper treatment of that regime would require an effective theory including the relevant mesonic degrees of freedom and is therefore beyond the scope of the present work.}}

\begin{figure}[H]
	\includegraphics[scale = 0.375]{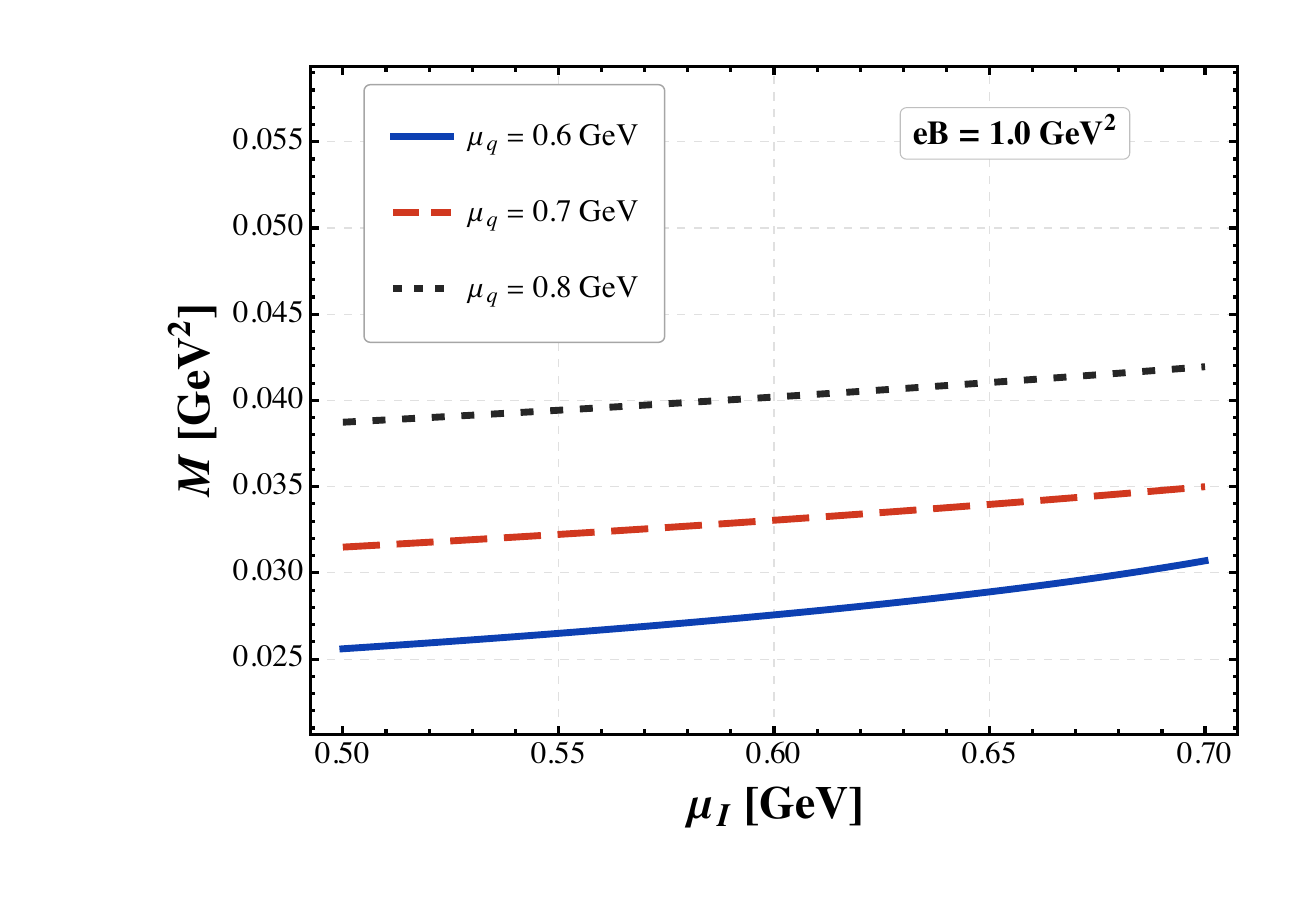}
	\includegraphics[scale = 0.375]{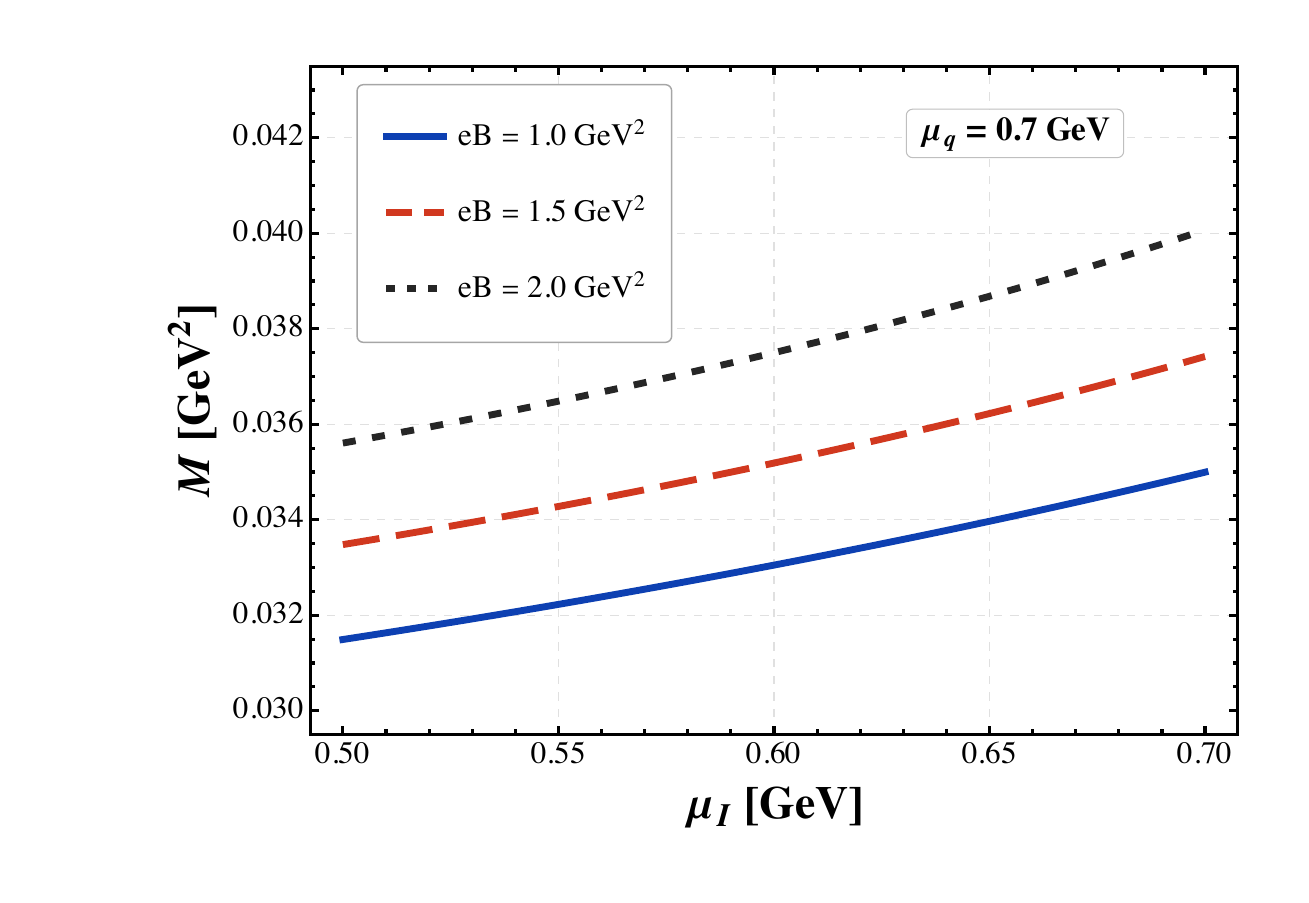} 
	\caption{Variation of magnetization with isospin chemical potential at different values of the quark chemical potentials at $|eB| = 1.0$ GeV$^2$ (left panel). Variation of the magnetization with isospin chemical potential at different strengths of magnetic field at $\mu_q = 0.7$ GeV (right panel). The plots have been obtained for $\Lambda = 2\mu$. }
	\label{magnetization}
\end{figure}

\section{Conclusions and outlook}
\label{conc}
In this work, we have investigated the thermodynamics of cold and dense quark matter at finite isospin density and zero temperature in a strong magnetic field. We have systematically employed the resummed perturbative scheme for hard dense loops, which includes medium effects such as screening and damping of soft modes. Our study shows that pressure increases with both isospin and quark chemical potentials. The pressure also increases with the magnetic field. We also computed the magnetization of cold quark matter, which is found to be positive, confirming the medium's paramagnetic nature. Our study provides insight into systems with isospin imbalance, such as the interiors of neutron stars and other compact astrophysical objects, where the densities of the $u$ and $d$ quarks differ. The results obtained at isospin chemical potential may have important implications for the mass–radius relation and stability of compact stars. {Note that the magnetic field used in the numerical analysis corresponds to the ultra-strong field regime. Our results should therefore be viewed as strong-field HDLpt estimates rather than as a direct description of all possible magnetic-field strengths in compact stars. Corrections from higher Landau levels are expected to become important when $\mu_f^2/(2|q_fB|)$ is no longer small.} Overall,  future investigations may include the effects of a nonzero quark mass on the thermodynamic properties of cold quark matter. The quark and gluon structure functions calculated herein can also be used to study the refractive index of cold quark matter, which we hope to study in the near future.

\section*{ACKNOWLEDGEMENTS}
S. A. K. is thankful to Integral University for providing the necessary facilities for research.

\section*{DATA AVAILABILITY}
The data supporting this study’s findings are available within the article.

\appendix
\section{Dense sum integrals}
\label{App-A}
 
Gorda et al.  \cite{Gorda:2022yex} showed that by taking the $T\to 0$ limit of the sum integrals calculated at finite $T$ and $\mu$ is not the same as simply putting $T=0$. He also emphasized that the $T \to 0$ limit of the finite $T$ and $\mu$ sum integrals is the correct way to obtain the dense sum integrals.
%At present, we have sum integrals with different powers of temporal, spatial, and four-momentum, respectively. 
The general structure of fermionic sum integrals at finite $T$ and $\mu$ is given by
\bea
&&\sumintof  \frac{ p_0^{2\beta} p^{2\omega} }{ P^{2\alpha} }  \nn &=& \frac{ T ~~\Gamma(\alpha - \omega - d/2) \Gamma(d/2 + \omega) }{(4\pi)^{d/2} \Gamma(\alpha) \Gamma(d/2) (2\pi T)^{2\alpha - 2\beta - d - 2\omega}} \big\{ \zeta(2\alpha - 2\beta - 2\omega - d, 1/2 + i\mu_f/ (2\pi T)) +  \zeta(2\alpha - 2\beta - 2\omega - d, 1/2 - i\mu_f/ (2\pi T))  \big\} , \nn 
\label{App-1}
\eea 
where $\zeta(z, q) = \sum_{n = 0}^{\infty} \frac{1}{(n + q)^z}$ refers to  Hurwitz zeta function. For our case, we need the $T/\mu_f \to 0$ limit of Eq. (\ref{App-1}), which corresponds to the dense limit of Eq. (\ref{App-1}). The dimensionally regularized form of the dense sum integrals reads as~\cite{Gorda:2022yex}
\bea  
\hspace{-0.55cm}\mathcal{I}_{\alpha\beta\omega}(\mu) &=&  \displaystyle{\lim_{T/\mu_f \to 0}}\sumintof  \frac{ p_0^{2\beta} p^{2\omega} }{ P^{2\alpha} }  \nn 
%%%%%%%%%%%%%%%%%%%%%%%%%%%%%%%%%%%%%%%%%%%%%%%%%%%%%%%%%%%%%%%%%%%
&=& \bigg( \frac{e^{\gamma_E} \Lambda^2}{4\pi}  \bigg)^\epsilon \frac{i\mu}{2\pi}\frac{ \Gamma(\alpha - \omega - d/2) \Gamma(d/2 + \omega) (i\mu)^{d + 2\omega - 2\alpha + 2\beta} }{ (4\pi)^{d/2} \Gamma(\alpha) \Gamma(d/2) \big( 1 + d + 2\omega - 2\alpha + 2\beta  \big)  } \Big\{ \big( -1\big)^{d + 2\omega - 2\alpha + 2\beta} -1 \Big\}. 
\label{SI-2}
\eea 
Using the above Eq. (\ref{SI-2}), we can compute  the various sum integrals needed to obtain the quark contribution to the free energy  (\ref{I_abc}) as
\bea
&&\mathcal{I}_{630}(\mu_f) = \bigg(\frac{\Lambda}{4\pi \mu_f}\bigg)^{2\epsilon}\frac{63}{4\pi\mu_f^4}\bigg[\frac{1}{512} + \frac{-1 + \log 256 + 4\log\pi + 2\psi(0,11/2)}{1024}\epsilon  \bigg] + \mathcal{O}(\epsilon^2) \nn
&&\mathcal{I}_{621}(\mu_f) = \bigg(\frac{\Lambda}{4\pi \mu}\bigg)^{2\epsilon}  \frac{1}{4\pi\mu_f^4} \bigg[ \frac{7}{512} + \frac{179 - 210\gamma_E + 420 \log 2\pi}{15360}\epsilon  \bigg] + \mathcal{O}(\epsilon^2)  \nn
&&\mathcal{I}_{612}(\mu_f) = \bigg(\frac{\Lambda}{4\pi \mu}\bigg)^{2\epsilon} \frac{1}{4\pi\mu_f^4}\bigg[\frac{3}{512} - \frac{3(1 + 10\gamma_E - 20\log 2\pi)}{5120}\epsilon \bigg] + \mathcal{O}(\epsilon^2) \nn
&&\mathcal{I}_{603}(\mu_f) = \bigg(\frac{\Lambda}{4\pi \mu}\bigg)^{2\epsilon} \frac{3}{4\pi\mu_f^4} \bigg[ \frac{1}{512} - \frac{ 9 - 10\log 4 }{512}\epsilon  \bigg] + \mathcal{O}(\epsilon^2)   \nn 
&&\mathcal{I}_{210}(\mu_f) = \bigg(\frac{\Lambda}{4\pi \mu_f}\bigg)^{2\epsilon} \frac{1}{4\pi}\bigg[ \frac{1}{2\epsilon} + 1 - \frac{\gamma_E}{2} + \log 2\pi  \bigg] + \mathcal{O}(\epsilon) \nn
&&\mathcal{I}_{201}(\mu_f) =  \bigg(\frac{\Lambda}{4\pi \mu_f}\bigg)^{2\epsilon}\frac{1}{4\pi }\bigg[\frac{1}{2\epsilon}  -1 - \frac{\gamma_E}{2} + \log 2\pi\bigg] + \mathcal{O}(\epsilon) \nn
&&\mathcal{I}_{420}(\mu_f) = - \bigg(\frac{\Lambda}{4\pi \mu_f}\bigg)^{2\epsilon}\frac{1}{4\pi\mu_f^2}\bigg[\frac{5}{16} + \frac{5\big(-1 + \log 16 + 2\log\pi + \psi(0,7/2)\big)}{16}\epsilon \bigg] + \mathcal{O}(\epsilon^2)\nn
&&\mathcal{I}_{402}(\mu_f) = - \bigg(\frac{\Lambda}{4\pi \mu_f}\bigg)^{2\epsilon}\frac{1}{4\pi \mu_f^2}\bigg[\frac{1}{16} + \frac{-5 - 3\gamma_E + \log 64 + 6 \log \pi}{48}\epsilon \bigg] + \mathcal{O}(\epsilon^2)  \nn 
&&\mathcal{I}_{411}(\mu_f) = - \bigg(\frac{\Lambda}{4\pi \mu_f}\bigg)^{2\epsilon}\frac{1}{4\pi\mu_f^2} \bigg[ \frac{1}{16} + \frac{1 - 3\gamma_E + \log 64 + 6\log \pi}{48}\epsilon \bigg] + \mathcal{O}(\epsilon^2) ,
\label{SI-1}
\eea 
where  $\psi(z, q)$ is the the generalized polygamma function which is given by  $\psi(z,q) = e^{-\gamma z} \frac{\partial}{\partial z} \big[ e^{\gamma z} \frac{\zeta(z + 1, q)}{\Gamma(-z)}\big]$.

\end{document}